\documentclass[11pt]{article}
\usepackage{jheppub}

\pdfoutput=1
\usepackage{amsmath,amssymb}
\usepackage{epsfig}
\usepackage{hyperref}
\usepackage{color}
\usepackage{slashed}
\usepackage{placeins}

\usepackage{amsfonts}
\usepackage{graphicx, rotating}
\usepackage{epstopdf}
\usepackage{latexsym}
\usepackage{rotating}

\usepackage[normalem]{ulem}

\usepackage{subfig}

\usepackage[export]{adjustbox}

\usepackage[shortlabels]{enumitem}

\usepackage[font={small}]{caption}   

\definecolor{myred}{rgb}{0.7, 0, 0}
\definecolor{myblue}{rgb}{0, 0, 0.7}
\definecolor{mygreen}{rgb}{0.04, 0.7, 0.5}

\hypersetup{colorlinks,citecolor=myred,linkcolor=myblue,urlcolor=myblue,linktocpage=true}

 \def\be   {\begin{equation}}   \def\ee   {\end{equation}}
 \def\ba   {\begin{array}}      \def\ea   {\end{array}}
 \def\bea  {\begin{eqnarray}}   \def\eea  {\end{eqnarray}}
 \def\bean {\begin{eqnarray*}}  \def\eean {\end{eqnarray*}}
 
 \def\bry{\begin{array}}
 \def\ery{\end{array}}

\numberwithin{equation}{section}

\begin{document}

\begin{flushright}
\footnotesize
TUM-HEP 1543/24 \\
\end{flushright}
\color{black}

\title{
Minimal Electroweak Baryogenesis via Domain Walls
}

\author[a,b]{Jacopo Azzola,}

\author[a]{Oleksii Matsedonskyi,}

\author[a]{Andreas Weiler}

\affiliation[a]{Technische Universit\"at M\"unchen, Physik-Department, James-Franck-Strasse 1, 85748 Garching, Germany}

\affiliation[b]{Ludwig Maximilian Universit\"at, Physik-Department, Theresienstraße 37, 80333 München, Germany}

\emailAdd{jacopo.azzola@tum.de, oleksii.matsedonskyi@tum.de, andreas.weiler@tum.de}

\abstract{
The Standard Model extended by a real scalar singlet \( S \) with an approximate \( \mathbb{Z}_2 \) symmetry offers a minimal framework for realizing electroweak baryogenesis (EWBG) during a first-order electroweak phase transition. In this work, we explore a novel mechanism where spontaneous \( \mathbb{Z}_2 \) breaking enables EWBG via domain walls separating two distinct phases of the \( S \) field. These domain walls feature restored (or weakly broken) EW symmetry in their cores and sweep through space, generating the baryon asymmetry below the temperature of EW symmetry breaking.  
We identify the key conditions for the existence of EW-symmetric domain wall cores and analyze the dynamics required for wall propagation over sufficient spatial volumes. Additionally, we outline the CP-violating sources necessary for baryogenesis under different regimes of domain wall evolution. 
The parameter space accommodating this mechanism spans singlet masses from sub-eV to  \( 15 \, \text{GeV} \), accompanied by a non-vanishing mixing with the Higgs boson.
Unlike the standard realization of EWBG in the minimal singlet-extended SM, which is notoriously difficult to test, our scenario can be probed by a wide range of existing and upcoming experiments, including fifth force searches, rare meson decays, and EDM measurements.
}

\maketitle


\section{Introduction}

Electroweak baryogenesis (EWBG)~\cite{Shaposhnikov:1987tw,Cohen:1990it}\footnote{See e.g. \cite{Morrissey:2012db,Cline:2006ts} for reviews.} is a well-studied framework for explaining the observed matter-antimatter asymmetry. One of its key ingredients --- the B+L violating EW sphaleron processes ---  is provided by the Standard Model (SM), while the other necessary  conditions~\cite{Sakharov:1967dj} --- the out-of-equilibrium dynamics and CP violation --- must arise from new physics. Such new physics often introduces particles or interactions accessible to current or near-future experiments, making EWBG an attractive and testable scenario.

\begin{figure}[t]
\center
\includegraphics[width=0.47 \textwidth]{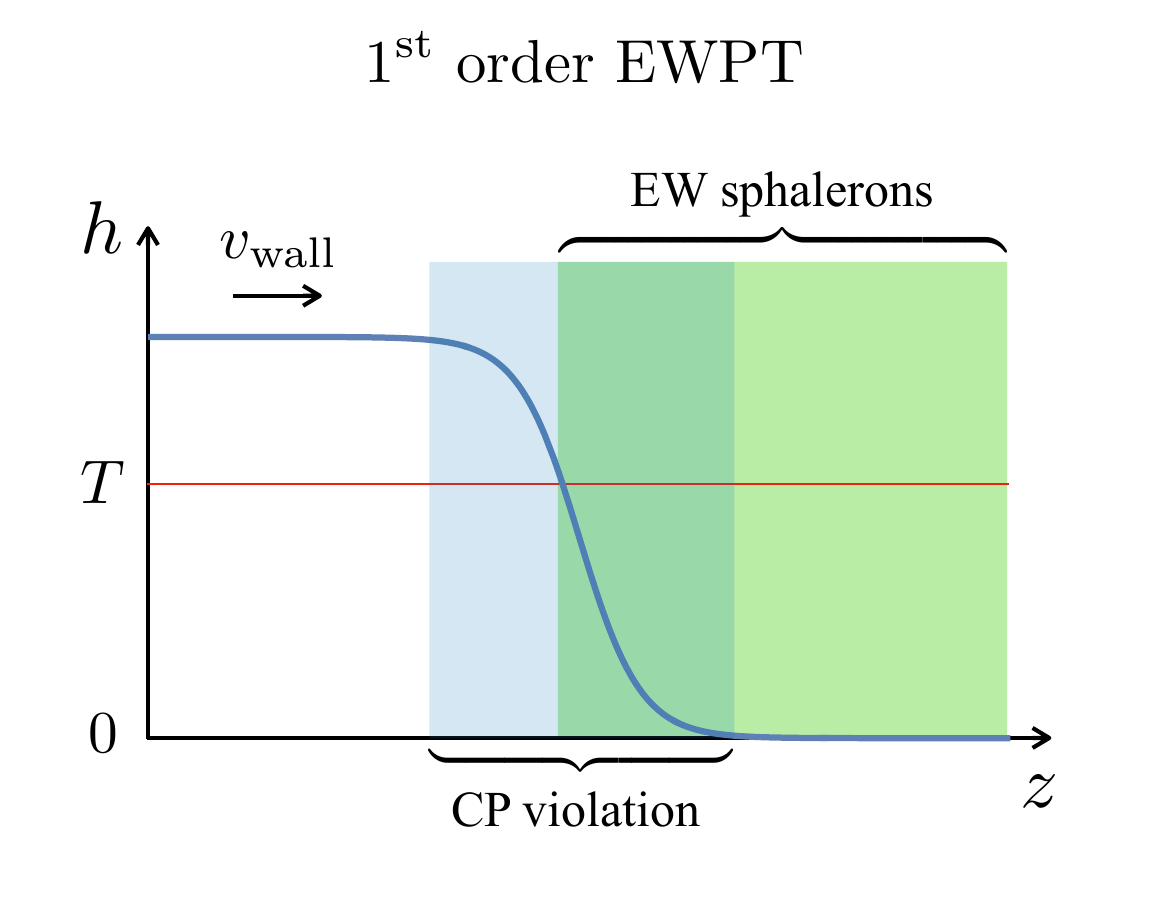}
\hspace{0.5cm}
\includegraphics[width=0.47 \textwidth]{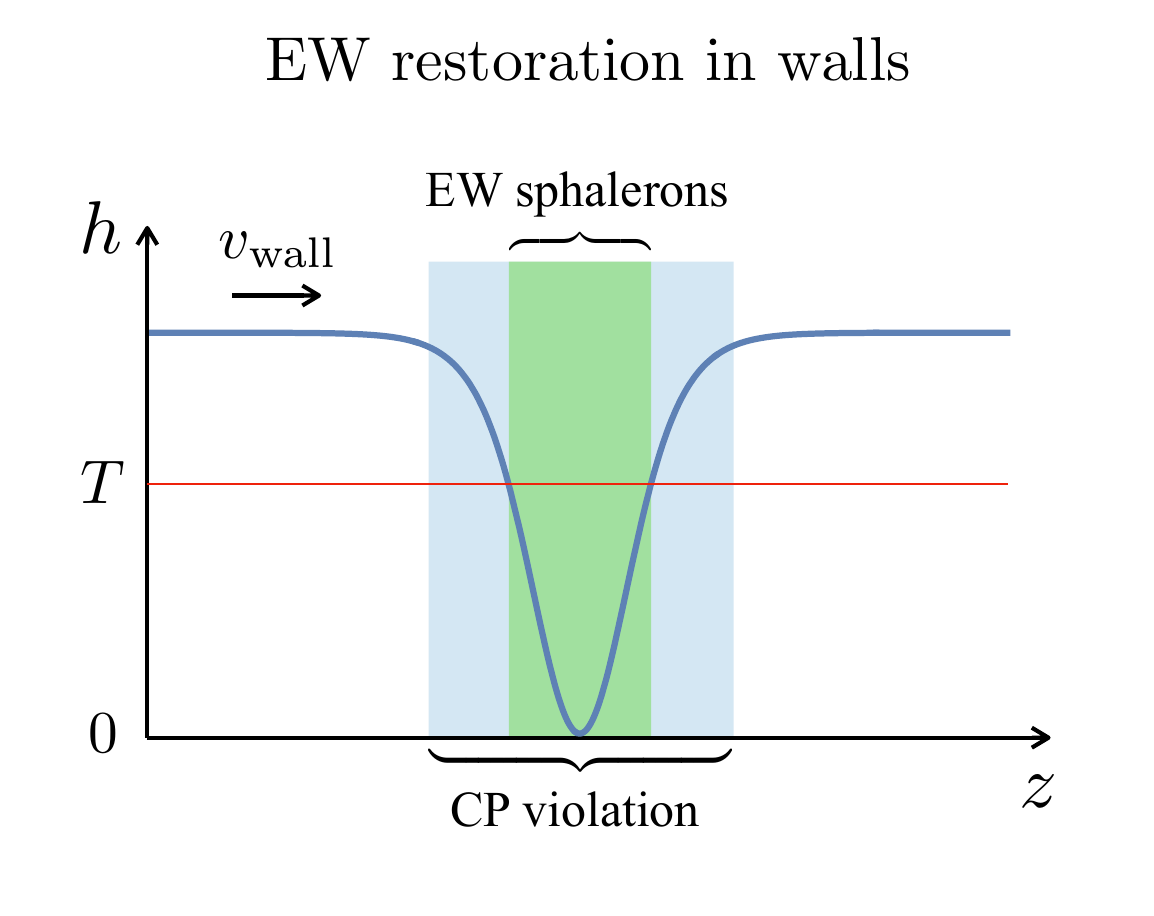}
\caption{{\bf Left:} 
Sketch of the relevant properties of the first-order electroweak phase transition. The bubble wall expands in the $z$ direction with a speed $v_{\text{wall}}$. The Higgs field's VEV transitions from \(h=0\) in the unbroken phase to \(h > T\) in the broken phase. EW sphalerons are effective in the $h<T$ regions. CP violation occurs within the bubble wall, where the scalar fields exhibit a non-zero gradient. This CP violation creates an excess of left-handed antiparticles in front of the wall, which gets processed by the EW sphalerons, leading to a baryon number asymmetry. {\bf Right:} Similar to the left panel, but for a domain wall separating two phases with equal Higgs VEVs (\(h > T\)) on either side. It is essential that in this case the core of the wall features restored (or weakly broken) EW symmetry, with \(h < T\). CP violation is localized within the wall, and the baryon number asymmetry is generated in the wall's core as it propagates through space. 
}
\label{fig:sketchintro}
\end{figure}

The conventional EWBG framework assumes baryogenesis to occur during a first-order EW phase transition. In this case, expanding bubble walls separating regions of broken and unbroken EW symmetry provide the required departure from equilibrium. Importantly, EW sphaleron processes are active outside the bubbles (where \( h/T < 1 \)) and inactive inside (if \( h/T > 1 \))~\cite{Morrissey:2012db,Cline:2006ts}. If the bubble walls exhibit CP-violating interactions with the surrounding plasma, a chiral asymmetry of left-handed particles and antiparticles forms across the walls. Sphalerons can then convert this asymmetry into a net baryon number. This process is schematically depicted in the left panel of Fig.~\ref{fig:sketchintro}.

The simplest model incorporating these ingredients is the real singlet extension of the SM with an approximate \(\mathbb{Z}_2\) symmetry~\cite{Espinosa:2011ax,Espinosa:2011eu,Ellis:2022lft,Carena:2019une,Beniwal:2017eik}. Depending on the parameter values, the potential barrier required for a first-order transition can arise either solely in the Higgs potential (with the singlet \(S\) remaining static) or in a combined Higgs-singlet potential (where both fields evolve during the transition). Moreover, if \(S\) varies during the transition, its coupling to the top quark can provide the necessary CP violation.

Interestingly, the EW phase transition is not the only cosmological process capable of generating spatial regions with broken and unbroken EW symmetry. At temperatures below the EW phase transition, when the universe predominantly resides in the broken EW phase, temporary symmetry restoration can occur within topological defects traversing the bulk of the broken phase. In particular, baryogenesis associated with domain walls was first proposed in Ref.~\cite{Brandenberger:1994mq} and further developed in e.g. Refs.~\cite{Abel:1995uc,Brandenberger:2005bx,Bai:2021xyf,Sassi:2024cyb,Schroder:2024gsi}~\footnote{See e.g. Ref.~\cite{Mariotti:2024eoh} for other domain walls-assisted baryon asymmetry production scenarios.}. 

The mechanism in this scenario, depicted in the right panel of Fig.~\ref{fig:sketchintro}, mirrors the case of a first-order phase transition in its reliance on EW sphalerons and in how CP violation is generated. However, the EW sphalerons are now confined to the cores of the domain walls. As the walls propagate through the broken phase, driven by surface tension and/or small potential energy differences across their boundaries, a baryon asymmetry is generated within their cores. We will discuss the details of this mechanism in the following sections.

\begin{figure}[t]
\center
\includegraphics[width=0.50 \textwidth]{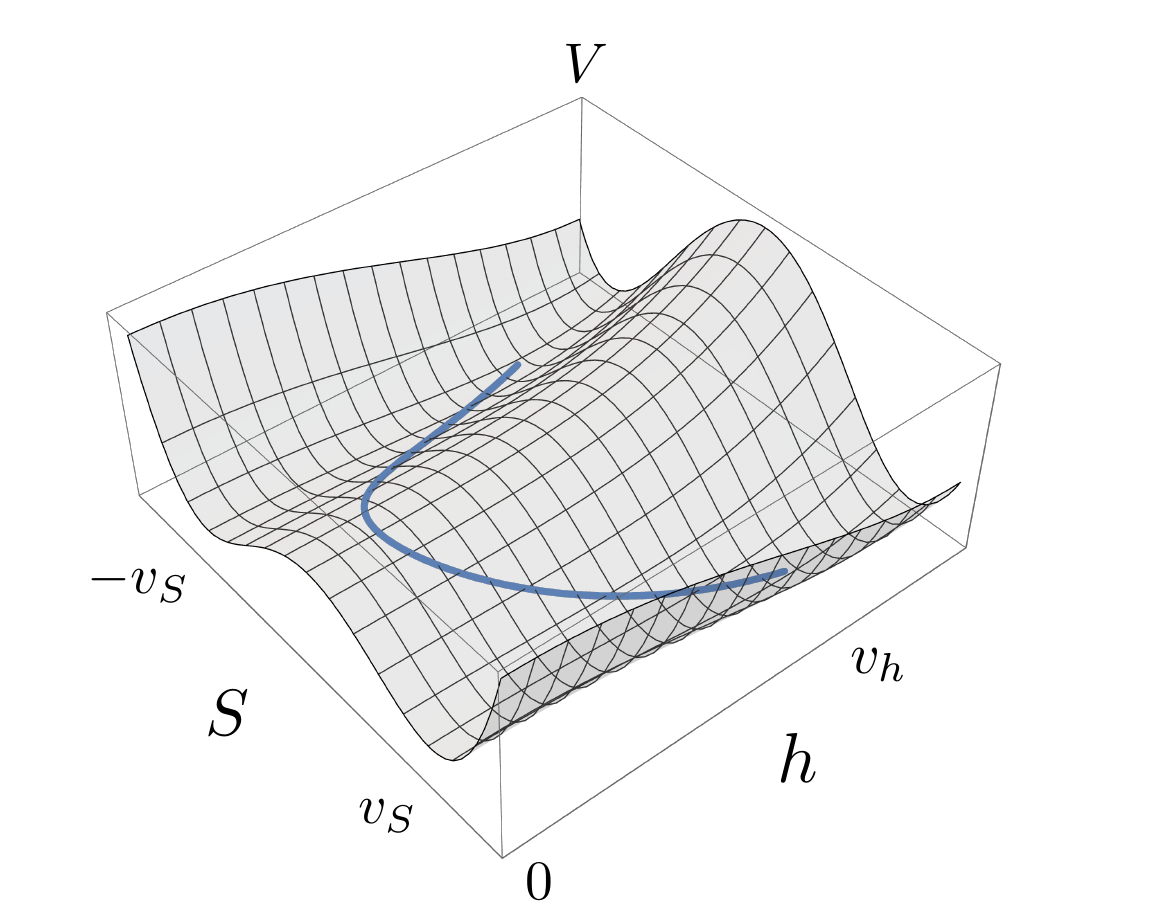}
\hspace{0.3cm}
\includegraphics[width=0.425 \textwidth]{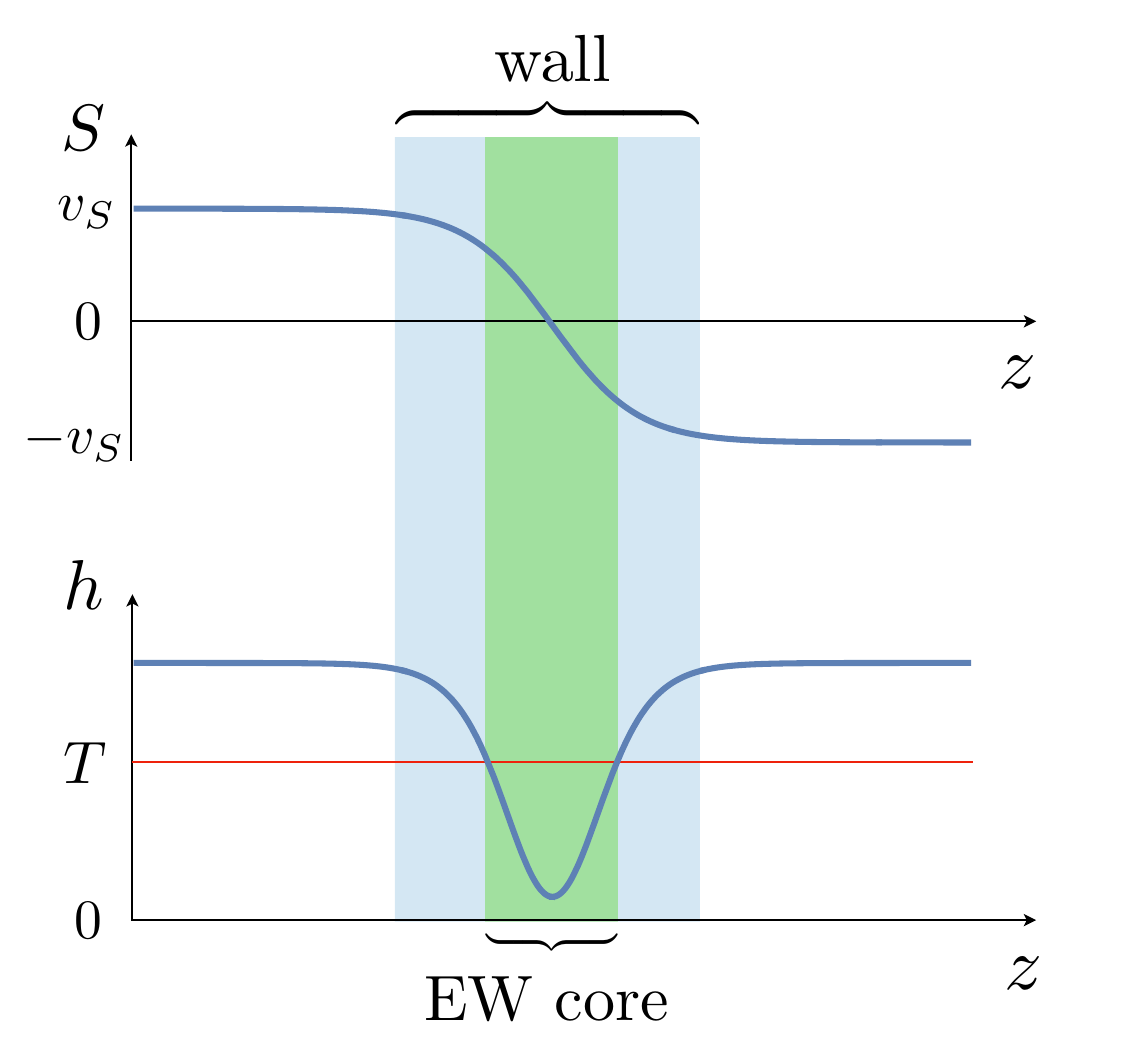}
\caption{{\bf Left:} example of a field trajectory (blue line) connecting two quasi-degenerate minima on opposite sides of the wall in the presence of $- \frac{|\lambda_{HS}|}{2} |H|^2 S^2$ contribution to the potential energy~$V$. {\bf Right:} $S$ and $h$ field profiles across the domain wall as functions of coordinate $z$, enabling the domain-wall mediated EWBG.}
\label{fig:sketchintro2}
\end{figure}

In this paper, we present, for the first time to our knowledge, how domain wall-mediated electroweak baryogenesis can be realized within the simple, approximately \(\mathbb{Z}_2\)-symmetric singlet extension of the Standard Model.

In the presence of $\mathbb{Z}_2$ symmetry, the vacuum manifold of the model supports two degenerate vacua, with opposite vacuum expectation values (VEVs) for the singlet $\langle S \rangle  = \pm v_S$.
Domain walls are then formed as field configurations  at the boundaries between these \(\pm v_S\) phases and smoothly interpolate between them.
As $S$ transitions from $-v_S$ in one domain to $+v_S$ in the other, it necessarily passes through $S=0$ in the core of the wall. In the presence of a negative cross-quartic coupling 
\be
V \supset - \frac{|\lambda_{HS}|} 2 |H|^2 S^2,
\ee
the Higgs mass squared will increase in the core, while its VEV will decrease, as illustrated in Fig.~\ref{fig:sketchintro2}. If the Higgs VEV drops below $T$, the EW sphalerons would get activated in the core.
Furthermore, we will introduce CP-violating interactions between the $S$ field and the SM particles in the plasma enabling the generation of local excesses of left-handed particles or antiparticles, so that the population of the latter can be depleted by the EW sphalerons.

The main differences of this scenario compared to the standard EWBG in $\mathbb{Z}_2$-symmetric extensions~\cite{Espinosa:2011ax,Espinosa:2011eu,Ellis:2022lft,Beniwal:2017eik} (except for~\cite{Carena:2019une})\footnote{$S$ domain walls are usually assumed to decay well before the EW symmetry breaking in the standard EWBG, however see~\cite{Blasi:2022woz,Agrawal:2023cgp,Blasi:2023rqi,Wei:2024qpy} for their possible effect on the EW phase transition if they are still present at that time.} are that 
\begin{itemize}
\item the $\mathbb{Z}_2$ symmetry is spontaneously broken in today's vacuum, leading to a non-negligible mixing between the Higgs and the singlet.
\item the EWBG-favoured singlet mass range extends down to much lower values, of the order $\sim10^{-5}$~eV. 
\end{itemize}
These two facts together lead to a broader range of potential probes including, besides collider experiments, astrophysical signatures and fifth force searches.

This paper is organized as follows. In Section~\ref{s:wallprofile} we discuss the conditions under which EW-symmetric cores form within the $S$ domain walls. Section~\ref{s:wallevol} explores different regimes of the large-scale evolution of the wall network, the volume swept by the walls, and their speed, which are crucial for the baryon asymmetry generation. The CP-violating sources are introduced and analysed in Section~\ref{s:bv} and we comment on possible UV completions in Section~\ref{sec:uv}. Section \ref{s:paramsp} examines experimental constraints on our scenario. Finally, conclusions are presented in Section~\ref{s:con}.

\section{EW Breaking/Restoration along the Wall Profile} \label{s:wallprofile}

In this section,  we explore a key component of our scenario: the possibility of local restoration of electroweak symmetry in the cores of domain walls, surrounded by the broken EW phase. To set the stage for this analysis, we first introduce the model's parametrization and discuss the leading-order temperature dependence of the relevant quantities.
We adopt the following Lagrangian for the singlet-extended SM
\be\label{eq:VHS}
{\cal L} = |D_\mu H|^2 + \frac 1 2 (\partial_\mu S)^2 - V_{\mathbb{Z}_2}(H,S) - V_{\slashed{\mathbb{Z}_2}}(H,S),
\ee
where the $\mathbb{Z}_2$-symmetric part of the potential is given by
\be\label{eq:vZ2}
V_{\mathbb{Z}_2}(H,S) = \mu_H^2 |H|^2 + \frac 1 2 \mu_S^2 S^2 + \lambda_H |H|^4 + \frac 1 4 \lambda_S S^4 + \frac 1 2 \lambda_{HS} |H|^2 S^2,
\ee
and $V_{\slashed{\mathbb{Z}_2}}(H,S)$ is a small $\mathbb{Z}_2$-breaking contribution necessary for the eventual decay of the domain walls. This contribution will be discussed in the next section. The remaining SM Higgs couplings are implicitly understood but suppressed here for clarity. In the following we will restrict our analysis to the case where the VEV of the Higgs doublet is aligned in one direction $H = (0,h/\sqrt 2)$ at all times, with $\langle h \rangle_{\text{today}} = v_{\text{SM}} = 246$~GeV. 
To derive analytic estimates in this section we will limit ourselves to the tree-level potential and leading thermal corrections. The effects of full one-loop quantum and thermal corrections are discussed in Appendices~~\ref{s:Vcorrections} and~\ref{s:numprofiles}.

The leading effect of the thermal corrections on the $h$ and $S$ potential amounts for replacing the mass parameters $\mu_{H,S}^2$ with the effective thermally-corrected quantities (see Appendix~\ref{s:Vcorrections}) 
\bea
\mu_H^2(T) 
&=& 
\mu_H^2 + T^2 \left\{ \frac {y_t^2}{4} + \frac{3g^2}{16} + \frac {g'^2}{16}  + \frac {\lambda_H}{2} + \frac{\lambda_{HS}}{24}  \right\}
\equiv 
\mu_H^2 + T^2 \left\{ c_T + \frac {\lambda_{HS}}{24}  \right\} \label{eq:muHT}
,\\ 
\mu_S^2(T) &=& 
\mu_S^2 
+T^2 \left\{ \frac{\lambda_{S}}{4} +  \frac{\lambda_{HS}}{6} \right\}, \label{eq:muST}
\eea
whereas the zero mode-induced next-to-leading thermal corrections $\propto T$  are subdominant and do not significantly impact the analysis.
The combined potential of the two fields~\eqref{eq:VHS} admits the possibility of having two quasi-degenerate minima forming a domain structure with $(h,S) \simeq \{(v_h, v_S), (v_h, -v_S)\}$ where
\bea
v_h^2 &=& \left({{-\mu_H^2 - \frac{\lambda_{HS}}{2} v_S^2  - T^2 \left\{ c_T + \frac {\lambda_{HS}}{24} \right\}}}\right)/{{\lambda_H}}, \label{eq:hvev} \\
v_S^2 &=& \left({{-\mu_S^2 - \frac{\lambda_{HS}}{2} v_h^2  - T^2 \left\{ \frac{\lambda_{S}}{4} +  \frac{\lambda_{HS}}{6} \right\}}}\right)/{{\lambda_S}}. \label{eq:Svev}
\eea
As we will discuss below, we do not find a sizeable variation of the $S$ VEV with temperature. Consequently, we use the notation \(v_S\) to refer to both the present-day value of \(S\) and its temperature-dependent minimum in the early universe. It is also worth noting that the thermal corrections induced by \(S\) are as shown in Eqs.~\eqref{eq:muHT} and \eqref{eq:muST} only if \(S\) thermalizes with the SM plasma. This depends on the value of the \(\lambda_{HS}\) coupling~\cite{Bernal:2018kcw}. However, given the relatively low impact of \(S\)-induced thermal corrections on our mechanism, we neglect this subtlety in our subsequent analysis. 

Next, we delineate the region of parameter space that enables the symmetry-breaking pattern described in the introduction. Specifically, we seek scenarios where the EW and \(\mathbb{Z}_2\) symmetries are broken in the bulk of space while sufficiently thick EW-restoring cores exist inside the domain walls.

\subsection{EW Core Width}

The main property of the walls that we are interested in is the presence of sufficiently extended EW-symmetry restoring cores (``EW cores'' in the following) with $h/T<1$, see Fig.~\ref{fig:sketchintro2}. 
These cores must have width, $l_{\text{core}}$, satisfying
\be\label{eq:lcore}
l_{\text{core}} > \left(\frac{g^2}{4 \pi} T\right)^{-1},
\ee 
in order to allow the EW sphalerons to efficiently process the excess of antiparticles~\cite{Brandenberger:1994mq}. 
Let us analyse the condition~\eqref{eq:lcore} in more detail. 

The domain walls correspond to field configurations connecting different $S$ VEVs, and their field profiles can be derived by solving the equations of motion 
\bea
h''(z) = \partial_h V  ,\;\; S''(z) = \partial_S V, \label{eq:wallprofile}
\eea
with the boundary conditions
\bea
h(z = \pm \infty) =  v_h ,\;\; S(z = \pm \infty) = \pm v_S,
\label{eq:bc}
\eea 
where $V$ is the overall scalar potential and $z$ is the coordinate perpendicular to the wall. 
The solutions correspond to infinite static plane walls, which is an adequate approximation when aiming to resolve the microscopic wall structure.
A simple solution for the $S$ field
\be\label{eq:sprof}
S(z) = v_S \tanh\left( \frac{m_{S}}{2} z \right),
\ee
with $m_S$ being the $S$ mass today, turns out to be a good approximation to the exact solution in most part of the relevant parameter space (see Appendix~\ref{s:numprofiles}). 
The typical overall width of the wall can then be estimated from Eq.~\eqref{eq:sprof} as
\be\label{eq:lwall}
l_{\text{wall}} \simeq 10/m_{S}. 
\ee

The next step is to relate the width of the $S$-wall with the width of the EW core. 
There are two qualitatively different regimes which one can consider here. 
To define them, let us first write down the second derivative of the Higgs potential at $h=0$
\be\label{eq:mhbare}
\partial_h^2 V (0,S) = \mu_H^2 + \frac 1 2 \lambda_{HS} S^2.
\ee
On one hand, this is simply proportional to the Higgs mass $m_h^2$ when $S=v_S$. On the other, when $S$ varies across the wall, the sign of the r.h.s. of Eq.~\eqref{eq:mhbare} signals whether the Higgs potential minimum is at zero or not.
In the tuned case with $|\lambda_{HS}| v_S^2,\, |\mu_H^2| \gg m_h^2$ one finds that
even a small decrease of $S$ causes the second derivative to turn positive and the Higgs potential minimum to move to zero\footnote{We do not discuss the Higgs mass tuning here, and just point out that there can be ways in which an apparently tuned Higgs mass can appear~\cite{Dvali:2003br,Graham:2015cka,Geller:2018xvz,TitoDAgnolo:2021nhd,Matsedonskyi:2023tca,Chattopadhyay:2024rha}.}. 
Although the gradient energy prefers more straight wall profiles with $h=const$, in this case the trajectory is strongly pushed by the scalar potential to the EW-restoring position $h=0$ along most of the wall width, which results in the EW core being almost of the same size as the overall width of the wall (see upper row of Fig.~\ref{fig:profilestudy} for illustration).

In the case of the untuned or weakly tuned Higgs mass we expect a more smooth $h(S)$ evolution, with $h$ gradually decreasing towards the center of the wall. To obtain precisely the core size in this case one needs to perform a numerical evaluation. However, generally one can expect that as long as the condition ensuring $h/T<1$ at the center of the wall is satisfied (see next sections), an order-one fraction of the wall width will have an EW core. 

The results of the numerical evaluation of the field profiles supporting these qualitative arguments are presented in Appendix~\ref{s:numprofiles}.
Assuming that the EW core takes an order-one fraction of the wall, we can use $l_{\text{wall}}$ of Eq.~\eqref{eq:lwall} for an estimate of the core size. Thus, the condition in Eq.~\eqref{eq:lcore} for the EW sphalerons to be efficient, results in a bound on the $S$ mass. Requiring this to hold for temperatures below  $\sim 100$~GeV, we obtain the requirement
\be
\boxed{m_{S} \lesssim \frac{10 g^2}{4\pi} T \lesssim 30 \,\text{GeV}}
\ee

\subsection{EW Breaking outside of Walls}\label{s:ewb}

The next important ingredient is the breaking of the EW symmetry in the bulk of space outside of the walls. In the Standard Model one expects the EW symmetry to get broken with $h/T>1$ at temperatures below about $130$~GeV. In our model, this conclusion could have been affected by the presence of the negative thermal correction to the Higgs mass $\propto \lambda_{HS} T^2$, increasing the breaking temperature. However a significant change of the breaking temperature is known to require special constructions, typically with a large multiplicity of BSM particles~\cite{Baldes:2018nel,Glioti:2018roy,Meade:2018saz,Matsedonskyi:2020kuy,Biekotter:2021ysx,Matsedonskyi:2021hti,vonHarling:2023dfl,Matsedonskyi:2022btb,Chang:2022psj,Aoki:2023lbz}, and is not  realized in our set-up.

The parameter space of our model instead features a region with an opposite effect -- decrease of the EW symmetry breaking temperature in the bulk of space. This is potentially harmful for our mechanism, since such an effect shortens the temperature range with the desired pattern of alternating EW symmetry breaking and restoration across the wall. To understand when this feature occurs let us start by writing down the mass of the light scalar singlet ($m_S \ll m_h$) at zero temperature. Using the approximate expression for the lighter mass eigenvalue $m_S^2 \simeq V''_{S} - (V''_{h,S})^2/m_h^2$ we find\footnote{The approximate expressions presented here hold up to corrections suppressed by additional powers of $\sin^2 \theta_{hS}$ and $m_S^2/m_h^2$.}
\be\label{eq:msdetails1}
m_S^2 
\,\simeq\, 
\mu_S^2 + \frac 1 2 \lambda_{HS} v_h^2 + 3 \lambda_S v_S^2 - \frac{\lambda_{HS}^2 v_h^2 v_S^2}{m_h^2}, 
\ee
from where, using the minimization condition $V'_S = 0$ to re-express the first two terms, together with $m_h^2 \simeq 2 \lambda_h v_h^2$, one derives
\be\label{eq:msdetails2}
m_S^2 
\,\simeq\, 2 \lambda_S v_S^2 - \frac{\lambda_{HS}^2 v_h^2 v_S^2}{m_h^2} 
\,\simeq\, 2 \lambda_S v_S^2 \left(1 - \frac{\lambda_{HS}^2}{4\lambda_H \lambda_S} \right).
\ee
Finally, we can use the approximate expression for the mass mixing between $h$ and $S$
\be\label{eq:sintheta}
|\sin \theta_{hS}| \simeq \frac{|\lambda_{HS}| v_h v_S}{m_h^2}
\ee
to rewrite the last term in~\eqref{eq:msdetails2} and obtain
\be\label{eq:msdetails3}
m_S^2 
\,\simeq\,  2 \lambda_S v_S^2 - \sin^2 \theta_{hS} \, m_h^2.
\ee 
 From the equations~\eqref{eq:msdetails2} and~\eqref{eq:msdetails3} it is clear that in the region with $\lambda_{HS}^2 \to 4\lambda_H \lambda_S$, or equivalently $|\sin \theta_{hS}| > m_S/m_h$, the physical $S$ mass is smaller than each of the two contributions it is composed of, meaning that the latter are finely tuned.  Generically it is unlikely to get in this tuned region\footnote{This also applies to numerical parameter space scans if performed in terms of the fundamental parameters $\mu_S^2, \lambda_{HS}, \lambda_S$.}, however there might exist mechanisms rendering the scalar in the tuned regime, such as those discussed in Refs.~\cite{Hook:2018jle,DiLuzio:2021pxd,Banerjee:2020kww}. 
 
Let us now find the dependence of the Higgs VEV on temperature in the tuned region using Eqs.~\eqref{eq:hvev}, \eqref{eq:Svev}  (assuming that $S$ has a non-vanishing VEV):
\bea
v_h^2 
=  
- \frac{\mu_H^2 - \frac{\lambda_{HS}}{2\lambda_S} \mu_S^2 }{\lambda_H - \frac{\lambda_{HS}^2}{4\lambda_S}}
-T^2 \left( \frac{  c_T  -\frac{\lambda_{HS}}{12} - \frac{\lambda_{HS}^2}{12 \lambda_S}}{\lambda_H - \frac{\lambda_{HS}^2}{4\lambda_S}} \right) 
=v_{\text{SM}}^2 - T^2 \left( \frac{  c_T  -\frac{\lambda_{HS}}{12} - \frac{\lambda_{HS}^2}{12 \lambda_S}}{\lambda_H - \frac{\lambda_{HS}^2}{4\lambda_S}} \right),
\eea
from which one sees that the Higgs VEV has an enhanced sensitivity to the thermal corrections when $\lambda_{HS}^2 \to 4\lambda_H \lambda_S$, leading to faster EW symmetry restoration\footnote{Note that given $\lambda_{HS}<0$, the tree-level potential becomes unbounded from below for $\lambda_{HS}^2 > 4\lambda_H \lambda_S$, which we demand not to happen.}.
The shrinking of the EWBG-viable temperature interval due to this effect does not happen as long as
\be\label{eq:boundEWB}
\boxed{|\sin \theta_{hS}| \lesssim \frac {m_S}{m_h}}
\ee
In the region, where the constraint of Eq.~\eqref{eq:boundEWB} is satisfied, the temperature of the EW symmetry breaking with $h/T>1$ in our model is close to that of the SM, $T_{\text{EWSB}} \simeq 130$~GeV.

\subsection{EW Restoration inside of Walls} \label{s:EWSR}

Let us now find what is necessary for the Higgs VEV to drop below $T$ in the wall. 
The change of the minimum of the Higgs potential, corresponding to $S$ variation from $S=\pm v_S$ outside of the wall to $S=0$ in the middle, is given by
\be\label{eq:deltah}
\Delta h^2 \,=\, h^2 (S=v_S) - h^2(S=0) \,\leq\,   -\frac{\lambda_{HS}}{2 \lambda_h} v_S^2,
\ee
so that a negative $\lambda_{HS}$ is needed for the minimum to move towards lower $h$ values.
Note however that $\Delta h^2$ of Eq.~\eqref{eq:deltah} is only an upper bound on the actual variation of the Higgs field, which does not follow precisely the minimum of its potential. Indeed, the change of $h$ value across the wall  costs extra gradient energy, and hence the actual energetically-optimal $h$ trajectory can have $h$ larger than the value minimizing the potential at a given $S$ (see left panel of Fig.~\ref{fig:sketchintro2} for an illustration).

Let us now recall that the alternating EW breaking/restoration requires 
\be\label{eq:hTh}
h(S=0,T) < T < h (S=v_S,T).
\ee
Neglecting the $T$-dependence in the Higgs field values we can then find an upper bound on the interval of temperatures satisfying Eq.~\eqref{eq:hTh}, which is given by $\Delta T^2 \equiv T_{\text{max}}^2 - T_{\text{min}}^2 < \Delta h^2$.
Combining this with~\eqref{eq:deltah} we get a useful constraint
\be\label{eq:deltaT}
\boxed{\Delta T^2 \lesssim \frac{|\lambda_{HS}|}{2 \lambda_h} v_S^2}
\ee
Dividing both sides of Eq.~\eqref{eq:deltaT} by $v_h^2$, and using the expression for the mixing angle~\eqref{eq:sintheta} we arrive at
\be
\frac{\Delta T^2}{v_h^2} \lesssim |\sin \theta_{hS}| \frac{v_S}{v_h},
\ee
which allows to conclude that in order to have a maximally possible interval of temperatures with the desired symmetry breaking pattern, $\Delta T^2 \sim (100\,\text{GeV})^2$, we would need
\be\label{eq:vsineq}
v_S \gtrsim v_h / |\sin \theta_{hS}| > v_h.
\ee

Finally, it is interesting to note that the condition of Eq.~\eqref{eq:deltaT}, when assuming $\Delta T \sim 100$~GeV, is equivalent to the requirement that the Higgs mass is not far from the tuned regime. See discussion below Eq.~\eqref{eq:mhbare} on this.

\subsection{Spontaneous $\mathbb{Z}_2$ Breaking}\label{s:Z2br}

Our starting assumption was the presence of $\mathbb{Z}_2$-breaking vacua with $S=\pm v_S$. 
We can now check how this $\mathbb{Z}_2$ breaking depends on temperature. The $T$-dependent VEV of the $S$ field is given by Eq.~\eqref{eq:Svev}. To analyse it, let us first note that only the thermal effects proportional to $\lambda_{HS}$ are important in $v_S$. Indeed, using Eq.~\eqref{eq:deltaT} with $\Delta T^2 \sim v_h^2$, together with $v_S^2 \sim m_S^2 / \lambda_S$ (follows from Eq.~\eqref{eq:msdetails3} assuming no tuning) we find 
\be
|\lambda_{HS}| > \frac{m_h^2}{v_S^2} \simeq \frac{m_h^2}{m_S^2} \lambda_S > \lambda_S.
\ee 
We now rewrite the $S$ VEV~\eqref{eq:Svev} taking this suppression of $\lambda_S$ into account, and singling out the $T=0$ value
\bea
v_S^2 &\simeq& v_S^2(T=0) 
+ \frac{|\lambda_{HS}|}{2\lambda_S} (v_h^2 - v_{\text{SM}}^2) 
+ \frac{|\lambda_{HS}|}{6\lambda_S} T^2. \label{eq:Svev2} 
\eea
One can now see that in the temperature interval $T=0\dots160$~GeV the value of $v_S$ decreases as $T$ grows. This is driven by the drop of the second term in Eq.~\eqref{eq:Svev2} $\propto v_h^2 - v_{\text{SM}}^2$, which is greater in absolute value than the growth of the third term $\propto T^2$ (recall that in SM, and similarly in our model, $v_h$ drops from $246$~GeV to zero while temperature grows from zero to $\sim 160$~GeV)

On the other hand, at temperatures $T\gtrsim 160$~GeV the Higgs VEV reaches zero, and the $S$ VEV evolution is driven by the thermal correction $\propto \lambda_{HS}$ alone, which makes $v_S$ grow with temperature. 

Therefore the minimum of the $S$ VEV is reached between the two regimes. 
Let us estimate whether it can vanish, while satisfying the previously derived conditions for the correct wall profile. For that to be true, the drop of the second term in Eq.~\eqref{eq:Svev2} happening when $T$ grows has to exceed at least the zero-temperature value $v_S^2(T=0)$, so that
\be\label{eq:z2sr}
{|\lambda_{HS}|} v_{\text{SM}}^2 \gtrsim 2 \lambda_S v_S^2(T=0) \simeq m_S^2,
\ee
where in the last step we assumed the untuned relation between the $S$ mass and VEV, see Eq.~\eqref{eq:msdetails2}.
On the other hand, from the no-tuning condition $\sin^2 \theta_{hS} < m_S^2/m_h^2$ together with the expression for the mixing angle~\eqref{eq:sintheta} we obtain 
\be\label{eq:noz2sr}
|\lambda_{HS}| v_{\text{SM}}^2 < m_S^2 \frac{m_h^2}{v_S^2} \frac{1}{|\lambda_{HS}|} <  m_S^2 \frac{m_h^2}{2\lambda_H \Delta T^2},
\ee
where we imposed Eq.~\eqref{eq:deltaT} in the last step. 
For sizeable temperature intervals of order $100$~GeV the inequalities~\eqref{eq:z2sr} and~\eqref{eq:noz2sr} are only compatible when both are close to saturation. 
However, given that they are only order of magnitude estimates, the final answer here requires a numerical evaluation, which did not give us any parameter space points where the $\mathbb{Z}_2$ symmetry is restored at some temperature. 

The restoration of the $\mathbb{Z}_2$ symmetry at some prior point in time however has a paramount importance for our scenario because it allows to naturally produce different $S$ domains after $\mathbb{Z}_2$ gets spontaneously broken.
We will therefore assume that a UV completion of our model takes care of the $\mathbb{Z}_2$ symmetry restoration at some temperature above $T_{\text{EWSB}} \simeq 130$~GeV.   
As we will see in Section~\ref{s:bv},  CP-violating operators in our minimal set-up also limit its validity at higher energies. However, in this regard our model of baryogenesis is not different from the vanilla EWBG during the first-order EW phase transition in the singlet-extended SM~\cite{Espinosa:2011ax,Ellis:2022lft,Beniwal:2017eik}.

\subsection{Parameter Space} \label{ss:parameterspace}

Let us now list the remaining constraints on the parameter space imposed by the EWBG mechanism. First of all, in all of the preceding discussion we were implicitly assuming the mass of $S$ to be greater than the Hubble parameter, ensuring negligible Hubble friction, and a clear separation between different domains. 
In order to have this constraint satisfied at all temperatures relevant for EWBG we need
\be\label{eq:msmin1}
m_S > H \simeq 1.7 \sqrt{g_*} \frac{T^2}{m_P} |_{ T\sim 100 \,\text{GeV}} \simeq 10^{-5}\,\text{eV}.
\ee
Additionally, one could demand that the $S$ field VEV does not reach the Planck scale $m_P$\footnote{This is not strictly necessary and can be consistently violated in some UV completions, see e.g.~\cite{Kaplan:2015fuy,Choi:2015fiu}.}. The minimal allowed mixing angle would then be given by
\be\label{eq:sinthmp}
|\sin \theta_{hS}| \simeq \frac{|\lambda_{HS}| v_h v_S}{m_h^2} > \frac{\Delta T^2}{v_h v_S}|_{\text{if} \,v_S<m_P} > 10^{-18},
\ee
where in the second step we used the condition~\eqref{eq:deltaT}, and in the last one we demanded $v_S < m_P$, and also took $\Delta T^2\sim (100\, \text{GeV})^2$.  
Combining the condition~\eqref{eq:sinthmp} with the EW breaking condition~\eqref{eq:boundEWB} we then find another lower bound $m_S \gtrsim 10^{-6}$~eV.  

Both presented above constraints however turn out to be comparable or weaker than the bound on the domain wall energy density derived at the end of Section~\ref{s:wallevol}. 
In order to avoid the energy density of the universe to be dominated by the domain walls, one has to satisfy the constraint~\eqref{eq:rhodw1} whose approximate form reads
\be\label{eq:rhodw4}
|\sin \theta_{hS}| \gtrsim \sqrt{\frac{m_S}{m_P}} \frac{v_{\text{SM}}}{T_{\text{decay}}},
\ee
which poses a lower bound on the temperature of the domain walls decay $T_{\text{decay}}$, thus limiting the overall temperature range of EWBG. In particular, if $T_{\text{decay}} > 130$ GeV, the domain walls will have to decay even before a sufficient EW symmetry breaking occurs, making EWBG impossible.

\begin{figure}[t]
\center
\includegraphics[width=0.8 \textwidth]{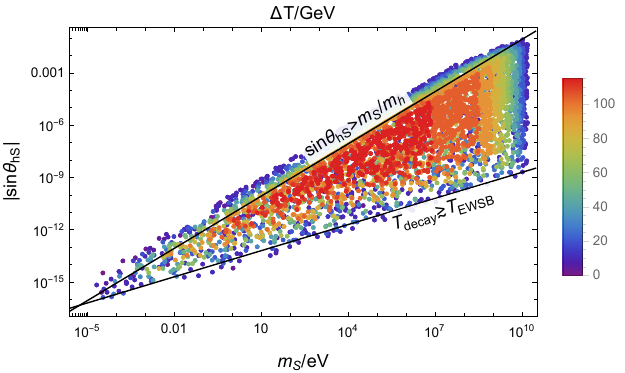}
\caption{Numerical parameter space scan. Colors show $\Delta T$ (with temperature scanned down to 1~GeV) during which the desired EW breaking pattern around the walls exists. Above the line with $\sin \theta \gtrsim m_S/m_h$ one has prolonged EW symmetry restoration outside of walls, lowering the upper boundary of $\Delta T$. For $m_S \gtrsim 15$~GeV the EW-restoring core inside of walls is too small. 
Approximately below the $T_{\text{decay}} \gtrsim T_{\text{EWSB}}$ line, given by Eq.~\eqref{eq:rhodw4}, the walls have to decay before EW symmetry breaking in order to avoid the energy density of the universe being dominated by them.}
\label{fig:paramspacescan}
\end{figure}

To confirm the analytic estimates of this section we present here the results of a numeric scan of the model parameter space.
We computed the thermal dependence of the field profiles using the tree-level zero-temperature potential with the leading-order thermal corrections. Furthermore, for this large scan we used a simplified way of computing the field profiles (``parabolic approximation''), discussed in the Appendix~\ref{s:numprofiles}. Such a simplified treatment allows for a larger scan and minimizes numerical errors. As also discussed in Appendix~\ref{s:numprofiles}, we have verified explicitly with a smaller number of parameter space points, that the results of this simplified treatment are sufficiently close to the ones obtained with full one-loop quantum and thermal corrections, and with a complete numerical solution for the field profiles. In particular, moderate order-one differences in $\Delta T$ (see below), when observed, were in favour of EWBG in the full analysis.

We scanned $m_S$, $\sin \theta_{hS}$, and $v_S$ uniformly on the logarithmic scale in the ranges $10^{-5}\, \text{eV} < m_S < 200\, \text{GeV}$, $10^{-20} < \sin \theta_{hS} < 1$, $10\, \text{GeV}< v_S < m_P$, and varied the temperature from 1~GeV to 200~GeV. 
The results of the scan are presented in Fig.~\ref{fig:paramspacescan} in terms of $m_S$, $\sin \theta_{hS}$ and the interval of temperatures $\Delta T = T_{\text{max}}-T_{\text{min}}$ during which the required pattern of EW symmetry breaking and restoration persists.
The maximal temperature $T_{\text{max}}$ is constrained by the temperature of the EW symmetry breaking with $h/T>1$ in the bulk of space. The minimal temperature $T_{\text{min}}$ is the greater of: the temperature at which the walls have to decay in order to prevent domination of the energy density; the temperature at which the EW core size is too small for EW sphalerons.
Scan of the physical mass $m_S$ rather than of the underlying parameters (e.g. $\mu_S^2$) allows to remove the statistical suppression of the tuned points above $\sin \theta_{hS} = m_S/m_h$. Hence the observed depletion of the number of points above this line happens entirely due to a physical effect, namely the drop of the maximal temperature $T_{\text{max}}$, as discussed in Sec.~\ref{s:ewb}. There is no viable points with $m_S \gtrsim 15$~GeV because of the decrease of the  EW core width, which then fails to satisfy the bound~\eqref{eq:lcore}. 
The lower boundary of the allowed region is defined by the constraint on the energy density stored in the walls~\eqref{eq:rhodw1}.

\section{Evolution of the Walls} \label{s:wallevol}

In our model, baryon production is mediated by domain walls and occurs exclusively in the regions swept by these defects. Thus, understanding the evolution of the domain walls is crucial.
As discussed in Section \ref{s:Z2br}, we assume that a  spontaneous $\mathbb{Z}_2$ breaking leads to the formation of the domain wall network at a temperature $T_{\text{d.w.}}$,  prior to electroweak symmetry breaking. 
As the temperature goes down to $T \sim 160$~GeV, the Higgs field will acquire a VEV in the bulk of space and the $h$-walls will appear inside of the $S$-walls. 

Initially, when the effect of the explicit breaking of the $\mathbb{Z}_2$ symmetry is negligible (we will quantify this condition later), the domain walls evolve by shrinking and smoothing to minimize surface area. Numerical studies of this evolution ~\cite{Press:1989yh,Leite:2011sc,Leite:2012vn,Garagounis:2002kt,Avelino:2005kn} found convergence to the so-called scaling regime, weakly dependent on the exact details of the model. The average comoving domain size $\xi_c$, defined as the ratio of comoving volume $V_c$ to the comoving surface $A_c$ of the walls inside of it, follows the scaling law  
\bea\label{eq:xic}
\xi_c & \equiv & \frac{V_c}{A_c} = \epsilon \eta,
\eea
where $\epsilon \simeq 0.5$ is found numerically and $\eta$ is the conformal time.
Also, the RMS velocity of the walls is found to be\footnote{This result neglects possible friction induced by the interaction of the walls with the SM plasma, see e.g.~\cite{Blasi:2022ayo}, which we will include in a future analysis.}
\bea\label{eq:vw}
\gamma v & \approx & 0.4. 
\eea

To estimate the fraction of comoving volume swept by the walls during the temperature interval $(T_i,T_f)$ where baryogenesis occurs, we consider the infinitesimal volume $dV_{\text{swept}}$ that is swept by a comoving surface $ds$ in a conformal time interval $d\eta$:
\be
dV_{\text{swept}} = v \, d\eta \, ds.
\ee 
Using the scaling law in Eq.~\eqref{eq:xic}, one then finds the volume fraction covered by the full wall network to be
\be
\frac{dV_{\text{swept}}}{V_{c}} 
= \frac{v \, d\eta}{\xi_c} \frac{d s}{A_c} 
= \frac{v}{\epsilon} \frac{d\eta}{\eta} \frac{d s}{A_c}.
\ee 
We now integrate this ratio in the temperature interval $(T_i, T_f)$,
 assuming $T_{\text{d.w.}}$ somewhat greater than $T_i$ to obtain
\be\label{eq:volume swept}
\frac{V_{\text{swept}}}{V_{c}} \simeq \frac{v}{\epsilon} \ln{\frac{T_i}{T_f}} + \frac{v}{\epsilon}\ln{\frac{T_{\text{d.w.}} - T_f}{T_{\text{d.w.}} - T_i}}.
\ee
Using as an example $T_i = 100$ GeV and $T_f = 1$ GeV we get $V_{\text{swept}}/V_{c} = \{6.1, 4.7, 4.6\}$ for $T_{\text{d.w.}} = \{130,10^3,10^6\}$~GeV. 
Note that these volume fractions exceed 1, which can be explained by the fact that the walls will typically pass through the same patch of space multiple times, if given enough time. This can potentially enhance the produced baryon asymmetry.  

\vspace{0.3cm}

An issue of the $\mathbb{Z}_2$-symmetric model is that domain walls redshift slower than matter and radiation, eventually dominating the universe energy density in contradiction with observations~\cite{Zeldovich:1974uw,Correia:2014kqa,Larsson:1996sp}. 
To avoid this situation we introduce a small potential energy difference $\delta V$ between the two $S$ vacua, thus breaking the $\mathbb{Z}_2$ explicitly. When this difference becomes large enough compared to the wall tension, the scaling regime ends and the higher-vacuum domains collapse. Concretely, the collapse occurs when the domain size grows to approximately
\be
\xi_{p}^{(\text{crit})} = \frac{\sigma}{\delta V},
\ee 
where $\xi_p \sim H^{-1}$ is now the physical correlation length of the wall network, defined as the ratio of the physical volume $V_p$ to the physical wall surface $A_p$, and $\sigma$ is the surface energy density of the walls.  
Requiring that the walls decay at temperature $T_{\text{decay}}$ we obtain a relation
\be\label{dVsigma}
\frac{\delta V}{\sigma} \sim 10^{-16}\text{ eV} 
\left(\frac{g_*(T_{\text{decay}})}{10}\right)^{1/2} \left(\frac{T_{\text{decay}}}{1 \text{ MeV} }\right)^2,
\ee
where $g_*$ is a number of relativistic degrees of freedom, and $T_{\text{decay}}$ should be higher than $\sim 1$~MeV to remove the walls before the BBN, which is necessary to minimize distortions in well-measured observables.\footnote{The  amount of $\mathbb{Z}_2$ breaking required by Eq.~\eqref{dVsigma} turns out to be \emph{lower} than what is expected to be generated by the Planck-suppressed operator of the lowest order, $\delta V \sim v_S^5/m_P$. This can be easily checked by taking $\sigma \sim m_S v_S^2$, and assuming the minimal allowed $v_S \sim v_{\text{SM}}/\sin \theta_{hS} \sim 10 v_{\text SM}$, and the maximal $m_S \sim 10$~GeV. Hence our scenario features a quality problem analogously to e.g. the QCD axion.}

We can verify that the required $\delta V$ is always small compared to the overall depth of the potential $V$, and thus does not contradict the assumption that both types of domains are equally populated from the start. For that, we demand
\be\label{eq:dVV}
\frac{\delta V}{V} 
= \frac{\delta V}{\sigma} \frac{\sigma}{V}
\simeq \frac{1}{\xi_{p}^{(\text{crit})}} \frac{1}{m_S}
\simeq  \frac{H(T_{\text{decay}})}{m_S}  
\ll 1,
\ee
where we approximated $\sigma = m_S^2 v_S^2 \times l_{\text{wall}} \simeq m_S v_S^2$, and $V = m_S^2 v_S^2$. The resulting bound $m_S \gg H(T_{\text{decay}})$ is the strongest when the walls decay soon after EW symmetry breaking, at $T_{\text{decay}} \sim 100$~GeV. Then the constraint~\eqref{eq:dVV} is similar to that of Eq.~\eqref{eq:msmin1}, giving $m_S \gg 10^{-5}$~eV, which is still weaker than the combination of bounds presented in Section~\ref{s:paramsp}.  

For each specific value of the surface energy density one can find the corresponding value of ${\delta V}$ needed to end the scaling regime at a given temperature $T_{\text{decay}}$ using Eq.~\eqref{dVsigma}. Other than setting the end time, this $\mathbb{Z}_2$ breaking does not affect sizeably the details of the wall evolution, as we have just checked. Hence we can trade ${\delta V}$ for $T_{\text{decay}}$ and treat the latter as a free parameter.  
Importantly, the collapsing domain walls in the $\delta V$-dominated regime are expected to experience a very low friction force~(see e.g. \cite{Espinosa:2011eu} and references therein) and therefore accelerate to $v \simeq 1$.

\vspace{0.3cm}

In addition to ensuring decay, we must avoid domain wall domination at any stage. Domain wall domination would lead to accelerated expansion, spoiling baryogenesis and potentially violating cosmological observations. 
As a result of this accelerated expansion, the walls are eventually expected to be frozen in the comoving coordinates~\cite{Avelino:2005kn}.    
Furthermore, even if the wall-dominated phase successfully ends everywhere, the energy density of the walls released into mildly-relativistic $S$ particles~\cite{Kawasaki:2014sqa} can either give an overabundance of dark matter (if $S$ is long-lived), or significantly dilute the previously produced baryon asymmetry (if $S$ efficiently decays into SM particles).
Overall,   
to prevent this regime 
we require that at all temperatures before the decay of the walls their energy density is subdominant:
\be\label{eq:rhodw1}
\frac{\sigma A_p}{\rho_{\text{rad}} V_p} = \frac{\sigma}{\rho_{\text{rad}} \xi_p} < 1,
\ee
where $\rho_{\text{rad}}$ is the radiation energy density. Using the estimate $\sigma \simeq m_S v_S^2$ we then find
\be\label{eq:rhodw2}
m_S v_S^2 \lesssim 0.2 \sqrt{g_*(T_{\text{decay}})} \, T_{\text{decay}}^2 m_P.
\ee
Furthermore, imposing the relation $v_S > v_{\text{SM}}/|\sin \theta_{hS}|$ of Eq.~\eqref{eq:vsineq} we obtain the bound
\be\label{eq:rhodw3}
\boxed{
|\sin \theta_{hS}| \gtrsim \sqrt{\frac{m_S}{m_P}} \frac{v_{\text{SM}}}{T_{\text{decay}}}
}
\ee    
where the temperature of the wall collapse $T_{\text{decay}}$ can not exceed the upper bound on the EWBG temperature $\sim130$~GeV. As we have shown in the previous section, this constraint provides the strongest lower bound on $|\sin \theta_{hS}|$.

In the next section, we discuss compatibility of the discussed regimes of the wall evolution with the baryon asymmetry generation.

\section{CP Violation and Baryon Asymmetry} \label{s:bv}

The domain walls with EW cores provide the mechanism allowing to process away local excesses of left-handed antiparticles, creating a net baryon number. In this section we will discuss the mechanisms allowing to create these local excesses in the first place, using CP-violating (CPV) interactions between the walls and the SM plasma. 
To this end, we will analyse two options for CPV interactions -- linear~(see e.g. \cite{Espinosa:2011eu}) and quadratic~(see e.g. \cite{Xie:2020bkl,Bai:2021xyf}) couplings between the singlet and the top quark 
\bea
{\cal L}_{\text{lin}} &=& - \frac {y_t}{\sqrt 2} h \bar t_L t_R \left(1 + i \frac {S}{f} \right) + h.c., \label{eq:cpvlin}\\ 
{\cal L}_{\text{quad}} &=& - \frac {y_t}{\sqrt 2} h \bar t_L t_R \left(1 + i \frac{S^2}{f^2} \right)  + h.c.,\label{eq:cpvquad}
\eea
where $y_t$ is the top quark Yukawa coupling, and $f$ is some scale suppressing the higher-order interactions. Such interactions can lead to asymmetric reflection of particles and antiparticles off the $S$ walls. The CPV-source entering the baryon number density evolution and responsible for the eventual baryon asymmetry is~\cite{Bruggisser:2017lhc} 
\be
S_{\text{CPV}} \propto \text{Im}[{m_t^\dagger}'' m_t] \propto 
\begin{cases}
\text{lin:}      & (h^2  S')' \\
\text{quad:}  & (h^2 S S')'
\end{cases} ,
\ee 
where $m_t$ is the effective $S$-dependent top quark mass defined by Eqs.~\eqref{eq:cpvlin} and~\eqref{eq:cpvquad}, and the derivatives are taken with respect to the $z$ coordinate directed perpendicular to the wall. An important difference between the two expressions presented above is that in the linear case the amount of CP and baryon asymmetry generated depends on the sign of the $S$ field or, more precisely, on whether the wall moves into the $+v_S$ domain or into the $-v_S$ one.

Note that $S_{\text{CPV}}$ is sensitive to the variation of $S/f$ across the wall, and hence a sizeable CP violation requires to have $f$ not too much greater than the range of $S$ variation, given by $v_S$. 

Let us now discuss the results~\eqref{eq:cpvlin},~\eqref{eq:cpvquad} in the context of different regimes of the wall evolution analysed in Section~\ref{s:wallevol}.
The scaling regime, occurring before the explicit $\mathbb{Z}_2$ breaking becomes relevant, seems most appropriate to host our mechanism because it is characterized by moderate wall velocities, of the order 0.4, which allow for efficient generation of the baryon number  (we assume that the velocity dependence of the efficiency of baryon production for the discussed mechanism is similar to the one of the charge transport within the standard EWBG during the first-order phase transition~\cite{Dorsch:2021ubz}).
At the same time, there is no preference to any specific type of domains in this regime. The domain walls are driven only by surface tension trying to smoothen out the local wall curvature and can convert both $+v_S \to -v_S$ and $-v_S \to +v_S$, depending on the local shape of the wall. 
Correspondingly, the sources of CP asymmetry which are odd under $S\to -S$ reflection will produce the CP (and B) asymmetry of opposite signs and on average will result in a vanishing or strongly suppressed baryon asymmetry.   
The quadratic coupling symmetric under $S\to - S$ and producing the asymmetry independently on the wall direction is instead free from this problem. 

The last stage of the wall evolution, driven by the explicit $\mathbb{Z}_2$ symmetry breaking contributions to the potential energy, instead features a wall movement in the well-defined direction (either $+v_S$ or $-v_S$ depending on which minimum is energetically preferred due to the $\mathbb{Z}_2$ breaking). It could therefore operate with the linear coupling. However, the wall collapse is generally expected to happen with the velocities $v\to 1$, which poses a challenge for the charge transport mechanism~\cite{Dorsch:2021ubz}. Nevertheless, a dedicated study in our specific set-up would be needed to address this question. 

We conclude that the scaling regime with the quadratic CPV coupling appears to be the optimal combination to produce the B asymmetry. 
We could expect that the rate of baryon number density production per domain wall crossing is similar to the one from the bubble wall passage in the case of the EWBG during the first-order phase transition, which is known to be able to generate a sufficient amount of baryon asymmetry. However, we leave a detailed study of this question to future work.

\section{Comments on UV completions}\label{sec:uv}

Although our analysis focuses on the phenomenological implications of the singlet-extended Standard Model, it is useful to comment on possible ultraviolet (UV) completions. In the region of parameter space where \( v_S \) is not significantly larger than the Higgs vacuum expectation value, this scenario could arise in models where both the Higgs and \( S \) emerge as Goldstone bosons of a spontaneously broken global symmetry near the TeV scale, such as in non-minimal composite Higgs models~\cite{Espinosa:2011eu}. 

Conversely, a significant portion of our model's parameter space involves an ultra-light singlet with a large VEV (\( v_S > v_{\text{SM}} / \sin \theta_{hS} \gg m_S \)) and a correspondingly small quartic coupling (\( \lambda_S \sim m_S^2 / v_S^2 \)). This configuration is characteristic of axion-like particle models. Without constructing a complete UV framework, we note that the potential of a pseudo-Goldstone boson coupled to the Higgs can take the following general form:

\[
V \supset \Lambda^4 \cos(S/f) + \kappa \Lambda^2 h^2 \cos(S/f + \phi),
\]
where $\phi$ is some constant phase. This potential exhibits similar properties to the potential in Eq.~\eqref{eq:vZ2}. In this framework, the mass of \( S \) is of the order of \( \Lambda^2 / f \), while \( v_S \sim f \). The suppression of the \( m_S / v_S \) ratio naturally arises from the scale hierarchy (\( \Lambda \ll f \)), where \( \Lambda \) may relate to the confinement scale of new strong interactions or explicit symmetry breaking, and \( f \) corresponds to the spontaneous symmetry-breaking scale producing the \( S \) Goldstone boson. 

Moreover, the coefficient \( \kappa \) required to match our model's parameters is typically \( \ll 1 \) and never exceeds a few, while \( \Lambda \) spans \( 10^2\text{--}10^6~\text{GeV} \). Additionally, incorporating the CP-violating operator \( i h \bar{t}_L t_R \cos(S/f) \) generates interactions analogous to those in Eq.~\eqref{eq:cpvquad}. 

A comprehensive exploration of UV completions is left for future work.

\section{Experimental Bounds} \label{s:paramsp}

\begin{figure}[t]
\center
\includegraphics[width=0.75 \textwidth]{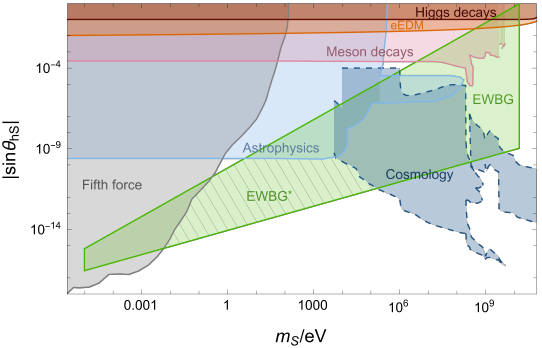}
\caption{Relevant experimental constraints on the parameter space. For simplicity, bounds are grouped into the categories discussed in the main text. The green area represents the most favourable for EWBG parameter space identified in Section~\ref{ss:parameterspace}. The hashed EWBG region with $m_S \lesssim 0.1$~MeV has a dark matter overproduction in the minimal model, see main text for details. }
\label{fig:exp}
\end{figure}

We will now summarize the main experimental constraints on our model for $S$ masses in the range $10^{-5}\,\text{eV} < m_S < 20\,\text{GeV}$, which are plotted in Fig.~\ref{fig:exp}.

For sub-eV masses, fifth-force experiments~\cite{Schlamminger:2007ht,Berge:2017ovy} pose the strongest constraints, excluding the gray area in Fig.~\ref{fig:exp}.
Next, astrophysics bounds apply up to the MeV scale. A new light scalar would affect heat and energy transport in astrophysical objects, modifying physical processes like star cooling or supernova explosions~\cite{Grifols:1988fv,Hardy:2016kme,Raffelt:2012sp,Bottaro:2023gep,Turner:1987by,Frieman:1987ui,Burrows:1988ah,Hardy:2024gwy,Caputo:2022mah,Diamond:2023cto}. Observational constraints of this type are shown in light blue in Figure~\ref{fig:exp}, together with the bound from XENON1T~\cite{Budnik:2019olh} sensitive to $S$ particles produced in the Sun.

Cosmological probes~\cite{Fradette:2018hhl,DEramo:2024lsk} can constrain the $S$ masses in the range $\left[10\,\text{keV},100\,\text{GeV}\right]$ and mixing down to $\sin{\theta_{hS}} \approx 10^{-16}$. These consider a production of the scalar particle in the early universe due to the mixing with the Higgs boson. Late decays during well-established phases of the universe evolution can then leave an imprint in cosmological observables like the CMB spectrum or affect BBN. Corresponding constraints on mass and mixing angle taken from Ref.~\cite{Fradette:2018hhl} are shown in darker blue. However, a more detailed investigation of applicability of these bounds to our specific set-up could be needed. 

Furthermore, the EWBG-viable region of the model parameter space below $m_S\simeq0.1$~MeV (hashed green in Fig.~\ref{fig:exp}) features an overproduction of dark matter~\cite{DEramo:2024lsk}. This happens because the decaying domain walls are expected to transfer an order-one fraction of their energy into mildly-relativistic $S$ particles~\cite{Kawasaki:2014sqa}, which quickly start red-shifting as matter and are stable on the cosmological timescales in that mass range. 
Considering how the energy density of the walls is constrained by EWBG, no available parameter space without the dark matter overproduction is left below $m_S\simeq0.1$~MeV.
While such a problem could in principle be avoided by adding extra ingredients in our set-up, we leave this for future work.

The MeV-GeV mass range is constrained by new physics searches at various accelerator experiments, sensitive to the $S$-mediated decays of $B$ and $K$ mesons, such as CHARM~\cite{CHARM:1985anb}, E949, Belle, BarBar and LHCb. We refer to Refs.~\cite{Banerjee:2020kww} and~\cite{Clarke:2013aya} for discussion of the relevant bounds, which are shown in pink in Fig.~\ref{fig:exp}.

The mixing between $h$ and $S$ fields produces a universal rescaling  of the Higgs couplings by a factor $\cos{\theta_{hS}}$. The strongest sensitivity~\cite{ATLAS:2022vkf,CMS:2022dwd} to such modifications comes from the coupling to $W$ and $Z$ vector bosons, $x_V > 0.936$ with $x_V = \cos{\theta_{hS}}$ being the ratio between the predicted Higgs coupling and the Standard Model one. This bound however turns out to be weaker than other discussed constraints, therefore it is not shown in Fig.~\ref{fig:exp}.

Moreover, the Higgs can decay into a pair of singlets. We show here in dark red the parameter space region excluded by the condition on the BSM Higgs Branching Ratio (BR)~\cite{ATLAS:2019nkf,Fuchs:2020cmm,Carena:2022yvx}: 
\be 
\text{BR}(h\rightarrow SS) = \frac{\Gamma_{h \rightarrow SS}}{\Gamma_{h,\text{total}}} \lesssim 0.2.
\ee

Importantly, the CP-violating interactions of the singlet with the top quark in Eqs.~\eqref{eq:cpvlin}, \eqref{eq:cpvquad} contribute at the two-loop level to the electron EDM constrained by the ACME collaboration~\cite{ACME:2018yjb}. The derivation of such a contribution is presented in Appendix~\ref{s:EDM}. For the quadratic CP-violating interaction of Eq.~\eqref{eq:cpvquad}, it takes the form
\be \label{eEDM d6}
	d_e^{\text{quad}} = \frac{2e}{3 \pi^2} \frac{\alpha G_F}{\sqrt{2} \pi} m_e \frac{v_\text{SM} v_S}{f^2} \frac{1}{1+v_S^4/f^4} \sin\theta_{hS} \cos\theta_{hS} \left(-g(z_{th}) + g(z_{tS})\right) \lesssim 2 \cdot 10^{-16} \text{GeV}^{-1},
\ee
where $z_{i,j} = m_i^2/m_j^2$ and the loop function $g(z)$ (see Appendix~\ref{s:EDM}) takes typical values of $1.4, 13.7$ for $z=m_t^2/m_h^2,m_t^2/m_e^2$. The above expression depends explicitly on the  scale $f$, which we take to be $2 v_S$ for definiteness. We can get a conservative bound in terms of $m_S$ and $\sin{\theta_{hS}}$ using the constraint $f/2 = v_S > v_{\text{SM}}/|\sin \theta_{hS}|$ of Eq.~\eqref{eq:vsineq}. This effectively results in the additional power of $\sin\theta_{hS}$ suppressing the EDM. Corresponding bound is shown in orange in Fig.~\ref{fig:exp}.
A simple numerical estimate of the constraint on the mixing can be obtained by taking an order-10 factor for the loop function, giving
\be 
\sin{\theta_{hS}} \lesssim 4 \cdot 10^{-2}.
\ee 

\begin{figure}[t]
\center
\includegraphics[width=0.75 \textwidth]{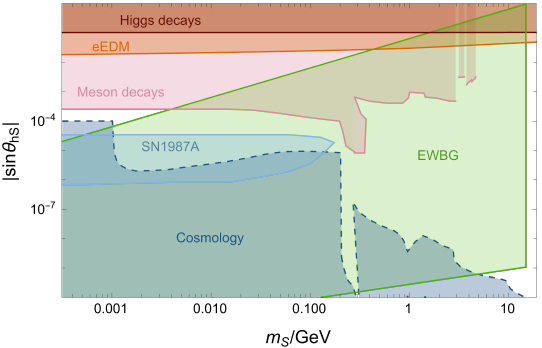}
\caption{Same as in Fig.~\ref{fig:exp}, but focused on the region without DM overproduction.}
\label{fig:expzoom}
\end{figure}

In summary, as can be seen from Fig.~\ref{fig:exp}, the presented bounds already exclude a sizeable fraction of the parameter space, leaving however a large portion of it unaffected, providing a target for future experimental searches. The best region of EWBG-viable parameter space without the dark matter overproduction problem is shown in Fig.~\ref{fig:expzoom}.

\section{Conclusions} \label{s:con}

Electroweak baryogenesis is an attractive possible explanation for the observed asymmetry between matter and antimatter, with a potential to be probed by the current and near-future experiments.
This scenario has been extensively studied in the recent years, with numerous concrete realizations proposed, see e.g.~\cite{Carena:2012np,vonHarling:2017yew,DeCurtis:2019rxl,Bruggisser:2022rdm,vonHarling:2023dfl}.
The most minimal one~\cite{Espinosa:2011ax,Espinosa:2011eu,Ellis:2022lft,Carena:2019une,Beniwal:2017eik} relies on the presence of a single new scalar degree of freedom with a $\mathbb{Z}_2$ symmetry, coupled to the Higgs and the top quark, and providing the first-order EW phase transition. 
However, this minimal model is notoriously difficult to test experimentally, as the $\mathbb{Z}_2$ symmetry not only reduces the model's parameter space to three variables, but also suppresses observable signatures by preventing mixing between the Higgs boson and the new scalar.

In this work, we proposed an alternative mechanism for baryogenesis within this minimal extension of the Standard Model. We demonstrated that the spontaneous breaking of the $\mathbb{Z}_2$ symmetry can produce domains separated by walls where the electroweak symmetry is restored. The interaction of these evolving domain walls with the SM particle plasma is analogous to the dynamics of bubble walls in conventional EWBG scenarios, enabling the generation of a net baryon asymmetry.

We have performed a study of the basic constraints on the parameter space of this model, concentrating on those ensuring the presence of alternating EW symmetry breaking and restoration across the walls.
We have also presented a preliminary study of CP violating sources in this set-up, while leaving a more comprehensive analysis for future work.

We found that the spontaneous breaking of $\mathbb{Z}_2$ symmetry persists till today, leading to a mass mixing between the Higgs boson and the new singlet, which can vary in a rather large range and thus be compatible with current experimental bounds. 
The allowed mass of the new singlet can be as large as $\sim 15$~GeV and can go down to $\sim10^{-2}$~eV, if the dark matter overproduction in the low-mass region is cured by an appropriate extension of the model. 
The predicted range of mass and mixing therefore can be probed by a wide range of different future experiments, from colliders to the fifth force searches.

Similarly to the realization of the usual EWBG in the singlet-extended SM, our model relies on higher-dimensional interactions to generate CP violation. The scale of these interactions defines the energy at which the model has to be UV completed. Furthermore, in this minimal framework, the UV completion must address the restoration of the  $\mathbb{Z}_2$ symmetry at high temperatures, a feature not achievable within the low-energy effective theory.

Taken together, these results highlight a rich avenue for exploring both the theoretical implications and experimental prospects of this minimal extension, which we will study in an upcoming work.

\vspace{1cm}
{\bf Acknowledgements}

We thank Majid Ekhterachian, Geraldine Servant, Stefan Stelzl, Mohamed Younes Sassi, Maximilian Bachmaier, Kai Bartnick and Michael Stadlbauer for useful discussions. 
The work of OM and AW has been partially supported by the Collaborative Research Center SFB1258, the Munich Institute for Astro- and Particle Physics (MIAPP), the Excellence Cluster ORIGINS, which is funded by the Deutsche Forschungsgemeinschaft (DFG, German Research Foundation) under Germany’s Excellence Strategy – EXC-2094-390783311. The work of JA has been partially supported by the Elitnetzwerk Bayern via the Elite Master Program in Theoretical and Mathematical Physics.

\appendix

\section{Effective Potential}\label{s:Vcorrections}

\subsection{Zero-Temperature}

We work in Landau gauge. The one-loop quantum corrections to the scalar potential are given by
\be
V_{\text{CW}} = \sum_i g_i \frac{(-1)^F}{64 \pi^2} m_i^4 \log m_i^2/\mu^2
\ee
where the sum runs over $i = \{t, W, Z, h, \chi, S\}$, $g_i = \{12, 6, 3, 1, 3, 1\}$, and $F=1(0)$ for fermions (bosons).
We will fix the counter-terms using the renormalization conditions
\bea
\partial_h V|_{v_h, v_S} &=& 0 \\
\partial_S V|_{v_h, v_S} &=& 0
\eea
where $V$ is the sum of tree-level potential, Coleman-Weinberg potential, and counterterms. $v_h, v_S$ are $T=0$ minima of the tree-level potential.

The next two conditions are derived by fixing the physical masses of the Higgs and the singlet. When computing the masses as the second derivatives of CW potential one encounters a divergence associated with massless Goldstone bosons. To fix it we account for the self-energy difference $\Delta \Pi_{ij}$ between the zero momentum transfer and the physical momentum $\mu_r$~\cite{Delaunay:2007wb}
\be
{\cal M}^2(\mu_r)_{ij} = \partial_i \partial_j V + \Delta \Pi_{ij} (\mu_r),
\ee 
where we only include the scalar-induced contributions to $\Delta \Pi_{ij}$, given by
\bea
\Delta \Pi_{hh}(\mu_r) &=& 
6 \lambda_H^2 v_h^2 \, \Delta \Pi(m_\chi, \mu_r) +  
18 \lambda_H^2 v_h^2 \, \Delta \Pi(m_h, \mu_r) +
(1/2) \lambda_{HS}^2 v_h^2 \, \Delta \Pi(m_S, \mu_r), \nonumber \\
\Delta \Pi_{hS}(\mu_r) &=& 
3 \lambda_H \lambda_{HS} v_h v_S \, \Delta \Pi(m_\chi, \mu_r) +  
3 \lambda_H \lambda_{HS} v_h v_S \, \Delta \Pi(m_h, \mu_r) +
3 \lambda_S \lambda_{HS} v_h v_S \, \Delta \Pi(m_S, \mu_r), \nonumber \\
\Delta \Pi_{SS}(\mu_r) &=& 
(3/2) \lambda_{HS}^2 v_S^2 \, \Delta \Pi(m_\chi, \mu_r) +  
(1/2) \lambda_{HS}^2 v_S^2 \, \Delta \Pi(m_h, \mu_r) +
18 \lambda_{S}^2 v_S^2 \, \Delta \Pi(m_S, \mu_r), \nonumber
\eea
where~\cite{Casas:1994us}
\bea
\Delta \Pi(m_i, \mu_r) &=& \frac{1}{16 \pi^2} \left(Z(m_i^2/\mu_r^2) - 2 \right), \nonumber \\
Z(x) &=& 
\begin{cases}
 x \geq 1/4     & 2 \sqrt{|1-4x|} \arctan(1/\sqrt{|1-4x|}) \\
 x \leq 1/4 & \sqrt{|1-4x|} \log\left( \frac{1+\sqrt{|1-4x|}}{1-\sqrt{|1-4x|}} \right)
\end{cases}.
\eea
The counterterms are then fixed such that eigenvalues of ${\cal M}^2(\mu_r)_{ij}$ are given by the physical masses when $\mu_r$ is set to each mass respectively.
Finally, we demand the one-loop corrections to the off-diagonal elements of ${\cal M}^2(m_h)_{ij}$ to vanish to fix the remaining renormalization condition.

\subsection{Thermal Corrections}

The one-loop thermal corrections are given by
\be\label{eq:VTgen}
\delta V (T) = \sum_i g_i \frac{(-1)^F T^4}{2 \pi^2} J_{F/B}(m_i^2/T^2),
\ee
where 
\be
J_{F/B}(x) = \int_0^{\infty} dk k^2 \log \left(1 \pm e^{-\sqrt{k^2 + x}}\right),
\ee
for fermions and bosons respectively.
We include the corrections from $i = \{t, W, Z, A, h, \chi, S\}$ with $g_i = \{12, 6, 3, 3, 1, 3, 1\}$. Note that due to thermal effects the three polarizations of each vector split in two transverse, which do not obtain sizeable corrections, and one longitudinal, which does. Furthermore, the masses of Goldstones $\chi$ also get split. These are discussed in detail below. 

In practice, to speed up the computation of the $J$ functions we used the approximation in terms of the modified Bessel functions of the second kind~\cite{Fowlie:2018eiu}
\be
J_{F/B}(y^2) = - y^2 \sum_{n=1}^{\infty} \frac{(\mp 1)^n}{n^2} \text{Re}[K_2(n y)], 
\ee
including three leading terms.
 
In high-temperature expansion, $m_i\ll T$, the thermal corrections read
\be
\delta V (T)|_{m_i \ll T} \simeq \sum_{i\in \text{\{bosons\}}} g_i \frac{T^2}{24} m_i^2 + \sum_{i\in \text{\{fermions\}}} g_i \frac{T^2}{48} m_i^2,
\ee
where we only show the leading terms relevant for us, in particular omitting the term $m^3T$ for bosons and field-independent terms.

When using the complete thermal potential~\eqref{eq:VTgen} in our numerical computations, 
we perform the daisy resummation by adding to the boson masses in Eq.~\eqref{eq:VTgen} the leading-order thermal corrections obtained in high-temperature expansion.
The thermally corrected masses of the longitudinal vector bosons read~\cite{Katz:2014bha}:
\bea
m_{W_L^{\pm}}^2(T) &=& 
g^2 \left\{\frac 1 4 h^2 + T^2 \left( \frac {2}{3} + \frac{3}{4} \theta_W + \frac{1}{4} \theta_t^{1/2} \theta_b^{1/2} + \frac{1}{24} (\theta_\chi^{1/2} + \theta_h^{1/2})^2 \right) \right\}, \nonumber \\
\left[m_{Z_L,A_L}^2(T)\right]_{1 1} &=&
 g^2 \left\{\frac 1 4 h^2 + T^2 \left( \frac {7}{8} + \frac{3}{4} \theta_W + \frac{1}{8} \theta_t  + \frac{1}{12} \theta_\chi \right) \right\}, \nonumber \\
\left[m_{Z_L,A_L}^2(T)\right]_{1 2} &=& 
- \frac{g g'}{4} h^2, \nonumber\\
\left[m_{Z_L,A_L}^2(T)\right]_{2 2} &=&
g'^2 \left\{\frac 1 4 h^2 + T^2 \left( \frac {109}{72} + \frac{51}{216} \theta_t  + \frac{18}{216} \theta_\chi \right) \right\}, \nonumber
\eea
while the scalars masses are
\bea
\left[m_{h,S}^2(T)\right]_{11} &=& 
\left\{ \mu_H^2 + \frac{\lambda_{HS}}{2} S^2 + 3 \lambda_H h^2 \right\} 
+ T^2 \left\{ \frac {g'^2}{16} + \frac{3g^2}{16} \theta_W + \frac {y_t^2}{4} \theta_t + \frac {\lambda_H}{4} \theta_h + \frac {\lambda_H}{4} \theta_\chi + \frac {\lambda_{HS}}{24} \theta_S \right\},  \nonumber \\
\left[m_{h,S}^2(T)\right]_{12} &=& 
\lambda_{HS} h S, \nonumber \\
\left[m_{h,S}^2(T)\right]_{22} &=& 
\left\{\mu_S^2 + \frac{\lambda_{HS}}{2} h^2 + 3 \lambda_S S^2\right\}
+T^2 \left\{ \frac{\lambda_{HS}}{24} \theta_h + \frac{\lambda_{HS}}{8} \theta_\chi + \frac{\lambda_{S}}{4} \theta_S \right\}, \nonumber \\
m_{\chi_{1,2}}^2(T) 
&=& \left\{ \mu_H^2 + \frac{\lambda_{HS}}{2} S^2 + \lambda_H h^2 \right\} 
+ T^2 \left\{ \frac {g'^2}{16} + \frac {3g^2}{16} \theta_W + \frac {y_t^2}{4} \theta_t^{1/2}\theta_b^{1/2} + \frac {\lambda_H}{12} \theta_h + \frac {5\lambda_H}{12} \theta_\chi + \frac {\lambda_{HS}}{24} \theta_S \right\},  \nonumber\\
m_{\chi_{3}}^2(T) 
&=& \left\{ \mu_H^2 + \frac{\lambda_{HS}}{2} S^2 + \lambda_H h^2 \right\} 
+ T^2 \left\{ \frac {g'^2}{16} + \frac {3g^2}{16} \theta_W + \frac {y_t^2}{4} \theta_t + \frac {\lambda_H}{12} \theta_h + \frac {5\lambda_H}{12} \theta_\chi + \frac {\lambda_{HS}}{24} \theta_S \right\}.  \nonumber
\eea
The masses are substituted in Eq.~\eqref{eq:VTgen} after diagonalization.
We defined $\theta_i = \theta(m_i,T)$ as suppression factors accounting for the deviation of the thermal corrections from the high-$T$ expansion. For example, a thermal correction to the Higgs mass from a bosonic particle species $i$ is given in the high-$T$ approximation by
\be
\delta m_h^2(T)  = \partial_h^2(\delta V(T)|_{m_i \ll T}) \simeq \frac{T^2}{24} \partial_h^2 m_i^2,
\ee
while a more precise expression, accounting for the Boltzmann suppression of $i$ states would read
\be\label{eq:deltamex}
\delta m_h^2(T)  = \partial_h^2(\delta V(T)) \simeq \frac{T^4}{2 \pi^2} \left\{ J_B' \frac{\partial_h^2 m_i^2}{T^2} + J_B'' \frac{(\partial_h m_i^2)^2}{T^4} \right\}.
\ee
We hence define $\theta$'s to reproduce the ratio between the exact and the approximate thermal corrections
\be
\theta(m_{\text{boson}},T) =  \frac{12}{\pi^2} J_B'(m_{\text{boson}}^2/T^2),\;\;
\theta(m_{\text{fermion}},T) =  -\frac{24}{\pi^2} J_F'(m_{\text{fermion}}^2/T^2),
\ee
where we only used the first term of Eq.~\eqref{eq:deltamex} which is typically dominant for $m\lesssim 2T$, while at larger $m$ the second term could dominate. However, at large $m$ both terms provide a fast exponential decay of the correction in any case. Moreover, the first term's ratio to the high-$T$ result is model-independent, in the sense that it only depends on the mass and temperature. 

We then fit $\theta$'s with simple exponentials allowing for fast numerical treatment
\be
\theta(m_{\text{boson}},T) \simeq e^{-0.9 |m_{\text{boson}}|/T},\;\; 
\theta(m_{\text{fermion}},T) \simeq e^{-0.4 (|m_{\text{fermion}}|/T)^{1.45}}.
\ee

When a diagram involves a loop with two states with different masses $m_{1,2}$ we weight it by the factor $\theta^{1/2}(m_1)\theta^{1/2}(m_2)$.  We set $\theta_b = 1$ since the bottom quark mass is sufficiently low, and $\theta_{h} = \theta_{S} = 1$ to simplify our numerical computations. For simplicity, in the expressions for the thermal masses presented above we took the same $\theta$ factors for $W_{1,2,3}$, all equal to $\theta(m_{W^\pm})$, and set $\theta = 1$ for the $B$ gauge field.

\section{Field Profiles}\label{s:numprofiles}

We discuss here numerical details of the computation of the wall profile for the two-field system.

The equations of motion~\eqref{eq:wallprofile} and \eqref{eq:bc} consist in a boundary value problem for a system of two non-linear second-order differential equations. To find a numerical solution we reformulate it as an initial value problem. Thus, the exact wall profile is found evolving an arbitrary initial field configuration via the time-dependent equations of motion with artificially introduced friction until the static solution is reached. 

Since this procedure is computationally demanding, the main parameter scan shown in Figure~\ref{fig:paramspacescan} was performed using a faster, albeit approximate, algorithm to find the wall profiles. We proceeded in the following way. First, we assumed the value of $h$ to be a function of  $S$, fixing the geometric shape of the wall trajectory in $(h,S)$ space. We adopted the following parabolic form for $h$
\be 
h(S,h_0) = \frac{v_h - h_0}{v_S^2} (S^2-v_S^2)+v_h,
\label{eq:parabolicansatz} 
\ee
where $h_0$ denotes the value of the $h$ field at the center of the $S$-wall and is found by minimizing the total energy of the two-field system within this subclass. With this approximation, we are only left with one effective equation of motion for $S$ that can be reduced to first-order via a Bogomolny method (see e.g.~\cite{Shifman_2012}) and then solved as a boundary value problem.
\begin{figure}
	\centering
	\includegraphics[width=.290\linewidth]{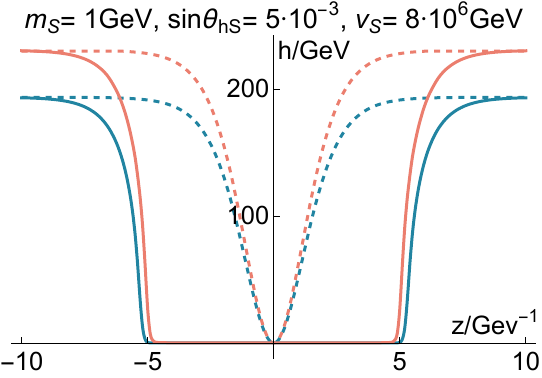}
	\includegraphics[width=.290\linewidth]{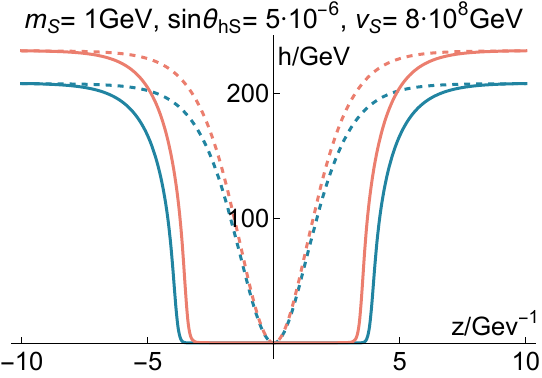}
	\includegraphics[width=.405\linewidth]{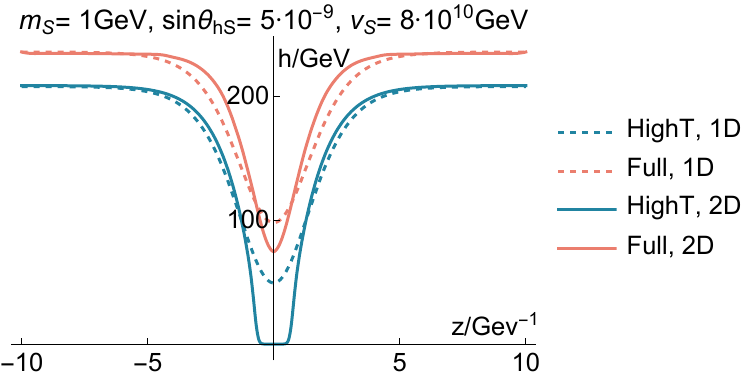}
	\includegraphics[width=.290\linewidth]{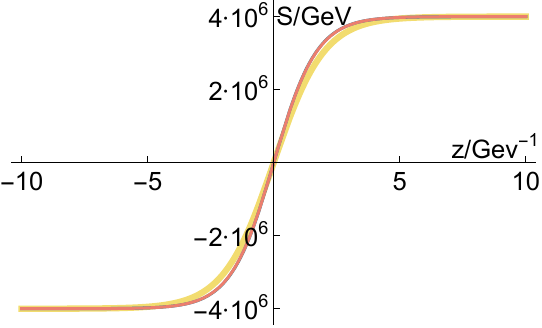}
	\includegraphics[width=.290\linewidth]{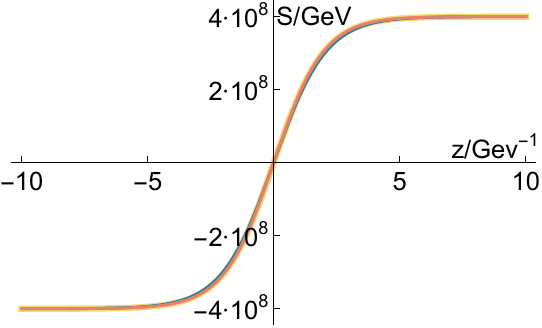}
	\includegraphics[width=.405\linewidth]{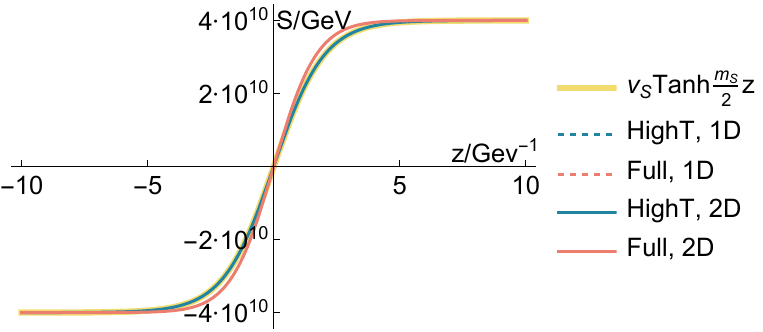}
	\caption{The first (second) row of plots shows the numerical results for the wall profile of the $h$ ($S$) field in different numerical approximations. ``HighT" refers to solutions with the tree-level potential and leading thermal corrections. ``Full" refers to solution with the one-loop potential and thermal corrections using the Bessel approximation for the $J$ functions, including daisy resummation and $\theta$ suppression (see Appendix~\ref{s:Vcorrections}). ``1D" (``2D") refers to the approximate (full) solution to the equations of motion as discussed in the main text.}
	\label{fig:profilestudy}
\end{figure}

The approximate solution was found to be very close to the exact one in the parameter space regions where the Higgs mass is not tuned. For tuned Higgs mass, the trajectory in $(h,S)$-space is much more angular than the parabolic ansatz~\eqref{eq:parabolicansatz}. Figure \ref{fig:profilestudy} compares the numerical wall profiles computed for tree-level and loop potentials with the two different numerical methods. The three columns correspond to parameter choices all having $m_S = 1$ GeV and varying $\sin{\theta_{hS}}$ and $v_S$ to change the level of tuning of the Higgs mass (more tuned to the left). The computation was done for $T$ = 70 GeV. 

The implications for our EWBG analysis are the following. First, the parabolic approximation is close to the full solution for untuned Higgs mass; elsewhere it always gives a lower estimate of the true EW core size. Second, as argued qualitatively in the main text, in untuned regions the EW core is an order-one fraction of the $S$-wall width; where the Higgs mass is tuned the core is usually as large as the $S$-wall width.

A second numerical challenge was posed by the stability of our computation procedure at low mixings. For this reason, we performed the main parameter scan using the tree-level potential and leading thermal corrections. 

To verify our approximations, we picked a set of points at the boundaries of the viable parameter space and compared the results for the EWBG temperature range computed with the two methods: using the approximate wall profiles with tree-level potential and leading thermal corrections, and using the exact wall profiles with full corrected potential. We found the temperature ranges to differ at most by an order-one factor, with the approximate value always being lower. This is the combination of two effects: first, the high-temperature expansion for the thermal potential overestimates the temperature corrections, resulting in a lower $v_h(T)$ at fixed $T$. Since $v_h(T)>T$ is needed to have broken electroweak symmetry outside the wall, this results in a lower upper bound for the temperature range in the approximate solution. On the other hand, the lower limit on the EWBG-viable temperature interval is usually given by the wall becoming too narrow for the sphalerons to be effective. As can be seen in Fig.~\ref{fig:profilestudy}, the $h$ profile usually stays closer to 0 for a much wider region in the full solution compared to the approximate one (at least in the case of a tuned Higgs mass). This directly results in an increase of the lower limit on the EWBG-viable temperature in the parabolic approximation. 

As a final remark, the bottom row in Figure \ref{fig:profilestudy} also shows as a yellow line the analytic solution given in Eq.~\eqref{eq:sprof}.

\section{Electron EDM} \label{s:EDM}

We derive here the contribution to the electric dipole moment of the electron discussed in Section~\ref{s:paramsp}, adapting to our model the result of Ref.~\cite{Keus:2017ioh}. In the course of this computation we will denote  the mass eigenstate fields by $\hat{h}$, $\hat{S}$ and  the interaction eigenstates by $h$, $S$.

We start with the dimension-5 CP-violating interaction from Eq.~\eqref{eq:cpvlin}
\be \label{CPV dim5 lagrangian}
	\mathcal{L} \supset - \frac{y_{t}}{\sqrt{2}} \bar{t}_Lt_R h \left(1 + i \frac{S}{f}\right) + h.c.
\ee
and expand it up to linear order in perturbations around the VEVs to find
\be
	\mathcal{L} \supset - \frac{y_{t}}{\sqrt{2}} v_\text{SM} \left(1+i\frac{v_S}{f}\right) \bar{t}_Lt_R - \frac{y_{t}}{\sqrt{2}} \left(1+i\frac{v_S}{f}\right) (h-v_{\text{SM}}) \bar{t}_L t_R - i \frac{y_{t}}{\sqrt{2}} \frac{v_\text{SM}}{f} (S-v_S) \bar{t}_L t_R
\ee
The first term gives mass to the top quark. To match the Standard Model value we impose 
\be
	\frac{y_{t}}{\sqrt{2}} v_\text{SM} \left(1+\frac{v_S^2}{f^2}\right)^{\frac{1}{2}}= m_{t},
\ee
and rotate away the phase, obtaining 
\be
 \mathcal{L} \supset - m_t \bar{t}_L t_R - m_t \frac{(h-v_\text{SM})}{v_\text{SM}} \bar{t}_Lt_R - \frac{im_t}{1+i v_S/f} \frac{(S-v_S)}{f} \bar{t}_Lt_R. 
\ee
Introducing now mass eigenstates $h-v_\text{SM} = c_{\theta} \hat{h} + s_{\theta} \hat{S}$ and $S-v_S = c_{\theta} \hat{S} - s_{\theta} \hat{h}$, with $c_{\theta}=\cos{\theta_{hS}}$ and $s_{\theta}=\sin{\theta_{hS}}$, we find 
\be \label{CPV dim5 lagrangian expanded}
	\mathcal{L} \supset - m_t \bar{t}_L t_R - m_t c_{\theta}\frac{\hat{h}}{v_\text{SM}} \bar{t}_Lt_R -m_t s_{\theta}\frac{\hat{S}}{v_\text{SM}} \bar{t}_Lt_R+ \frac{im_t}{1+i v_S/f} s_{\theta} \frac{\hat{h}}{f} \bar{t}_Lt_R -\frac{im_t}{1+i v_S/f} c_{\theta} \frac{\hat{S}}{f} \bar{t}_Lt_R.
\ee

\begin{figure}
	\center
	\includegraphics{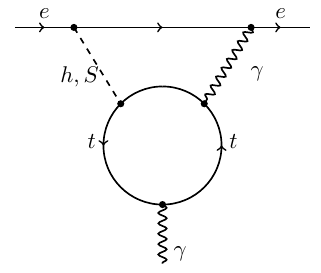}
	\caption{Barr-Zee diagram contributing to the electron EDM }
	\label{fig::BarrZee}
\end{figure}

The main contribution to the electron EDM comes from the 2-loop Barr-Zee diagram~\cite{Barr:1990vd} in Figure~\ref{fig::BarrZee}. The relevant result from \cite{Keus:2017ioh} reads:

\be
	d_{e,~f}^{\text{2-loop}}=
	\frac{e}{3\pi^2}
	\left(
	\frac{\alpha G_F v_{\text{SM}}^2}{\sqrt{2}\pi m_f}\right) \sum^n_{i=1}
	\biggl[\text{Re}(Y^{h_i}_{ee})\text{Im}(Y^{h_i}_{ff}) g(z_{fh_i})
	\biggr],
\ee
where subscript $f$ denotes the fermion species running in the loop (the top quark in our case), $h_i$ denotes $h$ and $S$ fields, $z_{AB}=m_A^2/m_B^2$, and $Y^{A}_{BC}$ denotes the Yukawa interaction between particles A,B and C. The function $g(z)$ is given by \cite{Keus:2017ioh}
\be  \label{g loop function}
	g(z)=\frac{1}{2}z\int_{0}^{1}dx\frac{1}{x(1-x)-z}\log\left(\frac{x(1-x)}{z}\right).
\ee 
We can read off the $Y$ coefficients for the top quark directly from Eq.~\eqref{CPV dim5 lagrangian expanded} and obtain analogously those for the electron from the Standard Model electron Yukawa. We get
\bea
		 \text{Re}(Y^{h}_{ee}) &=& - c_{\theta} \frac{m_e}{v_\text{SM}} \nonumber\\
		 \text{Re}(Y^{S}_{ee}) &=& - s_{\theta} \frac{m_e}{v_\text{SM}} \nonumber\\
		 \text{Im}(Y^{h}_{tt}) &=&  s_{\theta} \frac{m_t}{f} \text{Re}\left(\frac{1}{1+i v_S/f}\right) = s_{\theta} \frac{m_t}{f} \frac{1}{1+v_S^2/f^2} \nonumber\\
		  \text{Im}(Y^{S}_{tt}) &=& - c_{\theta} \frac{m_t}{f} \text{Re}\left(\frac{1}{1+i v_S/f}\right) = - c_{\theta} \frac{m_t}{f} \frac{1}{1+v_S^2/f^2}. \nonumber
\eea 
The final result for the electron EDM contribution is then
\be \label{eEDM d5}
	d_e^{\text{lin}} = \frac{e}{3 \pi^2} \frac{\alpha G_F}{\sqrt{2} \pi} m_e \frac{v_\text{SM}}{f} \frac{1}{1+v_S^2/f^2} \sin\theta_{hS} \cos\theta_{hS} \left(-g(z_{th}) + g(z_{tS})\right).
\ee 
We can generalize this result to the quadratic CP-violating interaction of Eq.~\eqref{eq:cpvquad}. The only differences are a factor of ${2 v_S}/{f}$ in the $S$-top vertex and the substitution $1/(1+v_S^2/f^2) \rightarrow 1/(1+v_S^4/f^4)$. The result is:
\be \label{eEDM d6}
	d_e^{\text{quad}} = \frac{2e}{3 \pi^2} \frac{\alpha G_F}{\sqrt{2} \pi} m_e \frac{v_\text{SM}}{f} \frac{v_S}{f} \frac{1}{1+v_S^4/f^4} \sin\theta_{hS} \cos\theta_{hS} \left(-g(z_{th}) + g(z_{tS})\right).
\ee
The presented results may be modified by an order-one factor for singlet masses below the electron mass, however in that case the parameter space is already better constrained by other types of observables, see Fig.~\ref{fig:exp}.

%
%

\bibliographystyle{JHEP} 
\bibliography{biblio} 

\end{document}